# Perturbative Lagrangian approach to gravitational instability

F.R. Bouchet[1], S. Colombi[1], E. Hivon[1], and R. Juszkiewicz[1,2]

[1] Institut d'Astrophysique de Paris, CNRS, 98 bis boulevard Arago, F-75014 Paris, France
[2] Copernicus Center, Warsaw.

**Abstract.** This paper deals with the time evolution in the matter era of perturbations in Friedman-Lemaître models with arbitrary density parameter $\Omega$, with either a zero cosmological constant, $\Lambda = 0$, or with a non-zero cosmological constant in a spatially flat Universe. Unlike the classical Eulerian approach where the density contrast is expanded in a perturbative series, this analysis relies instead on a perturbative expansion of particles trajectories in Lagrangian coordinates. This brings a number of advantages over the classical analysis. In particular, it enables the description of stronger density contrasts. Indeed the linear term is the famous Zel'dovich approximate solution (1970). This approach was initiated by Moutarde et al. (1991), generalized by Bouchet et al. (1992), and further developed by many others. We present here a systematic and detailed account of this approach. We give analytical results (or fits to numerical results) up to the third order (which is necessary to compute, for instance, the four point spatial correlation function or the corrections to the linear evolution of the two-point correlation function, as well as the secondary temperature anisotropies of the Cosmic Microwave Background). We then proceed to explore the link between the lagrangian description and statistical measures. We show in particular that Lagrangian perturbation theory provides a natural framework to compute the effect of redshift distortions, using the skewness of the density distribution function as an example. Finally, we show how well the second order theory does as compared to other approximations in the case of spherically symmetric perturbations. We also compare this second order approximation and Zel'dovich solution to N-body simulations in the description of large-scale structure formation starting from a power law ($n = -2$) power spectrum of Gaussian perturbation. We find that second order theory is both simple and powerful.

*Send offprint requests to*: F.R. Bouchet

## 1. Introduction

The observed large scale structures revealed by galaxy catalogs are usually supposed to have arisen due to gravitational instability acting on small initial perturbations in the context of an expanding Universe. The corresponding dynamics may in principle be described either from an Eulerian or Lagrangian point of view. With the notable exception of Zel'dovich (1970) approximate solution (and numerical simulations), analyses have up to very recently been largely dominated by the use of the Eulerian approach. Actually, most Eulerian studies concentrated on the perturbative regime, when density contrasts and velocities are small, and calculations are tractable. By principle, thus, such an approach is limited to the description of weak density contrasts. This is not the case of the Lagrangian approach, as is exemplified by the case of one-dimentional perturbations when Zel'dovich approximation is in fact exact up to the singularity, when the density contrast becomes infinite!

In order to contrast the two points of view, we start by recalling the classical Eulerian perturbative approach. Then we review some recent developments concerning the Lagrangian perturbative approach introduced by Moutarde et al. (1991). In this paper, as in Bouchet et al. (1992), we study directly the Newtonian perturbative approach, and extend this work in several ways. In the next section (§2), we establish notations and give some details of the derivation which was, due to lack of space, only sketched in the letter by Bouchet et al. (1992). In keeping with Zeldovich's spirit, we focus on the fastest growing modes of irrotational flows. We also provide quite accurate approximations to the third order solutions (which cannot be expressed exactly, for $\Omega \neq 1$, by using simple mathematical functions). Similarly we give expressions up to third order in the flat $\Lambda \neq 0$ case. In §3 we relate statistical indicators to the properties of the initial conditions. This is applied to the real space–redshift space mapping of low order statistical properties. Section 4 is devoted to a systematic comparison of the Lagrangian second order solution with other approximations, in the spherically symmetric case whose exact evolution is well-known. First and second order Lagrangian approximations are then compared to N-body simulations in the case of Gaussian initial conditions with a power-law power spectrum ($n = -2$) in §5. We conclude in §6 by a discussion.

### 1.1. Classical Eulerian Approach

Our aim here is to recall the basic steps of the perturbative Eulerian approach. The same conceptual steps are involved in the Lagrangian case. But the simple change of "small parameter" governing each of the two expansions (i.e. the density contrast for the Eulerian approach versus the displacement field for the lagrangian one) has far reaching consequences, in terms of the range of validity of each approach, and more generally in terms of the class of problems tractable analytically. A more detailed account of the Eulerian theory can be found for instance in Peebles (1980, hereafter LSS).

Consider perturbations which may be described in the Newtonian approximation. It is convenient to use comoving coordinates,

$$\boldsymbol{r} = a\boldsymbol{x}, \quad \text{and} \quad \boldsymbol{p} = ma^2 \frac{d\boldsymbol{x}}{dt},$$

where $m$ and $a$ stand respectively for the particles mass and the scale factor of the metrics of the Friedman-Lemaître background model. The motion and field (Poisson) equations then read

$$\frac{d\boldsymbol{p}}{dt} = -\nabla\phi, \quad \nabla^2\phi = 4\pi G a^2 \overline{\rho}\, \delta, \tag{1}$$

if $\delta \equiv \rho/\overline{\rho} - 1$, and $\overline{\rho}$ is the mean mass density.

The kinetic theory equation governing the evolution of the system then obtains by requiring *mass to be conserved*. If $f(\boldsymbol{x}, \boldsymbol{p}, t)$ stands for the probability density of finding at time $t$ a collisionless particle within the infinitesimal volume $d\boldsymbol{x}d\boldsymbol{p}$, Liouville theorem yields the Vlasov (or collisionless Boltzmann) equation

$$\partial_t f + \frac{\boldsymbol{p}}{ma^2} \cdot \nabla f - m\nabla\phi \cdot \partial_p f = 0,$$

where partial derivatives over $t$ are denoted by $\partial_t$. Taking velocity moments of this mean-field equation then leads to an infinite hierarchy. In particular, the macroscopic density, $\rho = ma^{-3}\int d\boldsymbol{p}\, f = \overline{\rho}(1+\delta)$, and macroscopic velocity, $\boldsymbol{v} = \int d\boldsymbol{p}\, \boldsymbol{p}\, f/(ma^2 \int d\boldsymbol{p}\, f)$, satisfy the conservation equation

$$a\overline{\rho}\partial_t \delta + \nabla(\rho \boldsymbol{v}) = 0 \tag{2}$$

and

$$\partial_{t^2}\delta + 2\frac{\partial_t a}{a}\delta = \frac{1}{a}\nabla \cdot [(1+\delta)\nabla\phi]$$
$$+ \frac{1}{m^2 a^4}\partial_\alpha \partial_\beta \left[(1+\delta)\langle p_\alpha p_\beta\rangle\right]. \tag{3}$$

In the pressureless case, $\langle(p_\alpha - mav_\alpha)(p_\beta - mav_\beta)\rangle = 0$; thus $\langle p_\alpha p_\beta\rangle/(m^2 a^2) = v_\alpha v_\beta$, which closes the hierarchy. Alternatively, one could start from the perfect fluid equation (LSS §5, p.47).

Perturbative solutions are then obtained by means of an iterative procedure. Let us write the density constrast as

$$\delta = \varepsilon\, \delta^{(1)} + \varepsilon^2 \delta^{(2)} + \varepsilon^3 \delta^{(3)} + \mathcal{O}(\varepsilon^4),$$

where $\varepsilon$ is just a book-keeping device, and $\varepsilon \ll 1$. First, keeping only the terms of order $\varepsilon$, one obtains

$$\partial_{t^2}\delta^{(1)} + 2\frac{\partial_t a}{a}\delta^{(1)} = 4\pi G a^2 \overline{\rho}\,\delta^{(1)},$$

which governs the linear evolution.

In order to solve this equation, one now has to specify the time variation of the scale factor. In the absence of a cosmological constant, and in the matter era when $\overline{\rho} \propto a^{-3}$, Friedman equation can be rewritten as follows

$$\left(\frac{\partial_t a}{a}\right)^2 = \frac{8\pi G\overline{\rho}}{3} - \frac{1}{a^2 R^2} = \frac{8\pi G\overline{\rho}}{3}\left[1 + (\Omega_0^{-1} - 1)\frac{a}{a_0}\right],$$

with the subscript 0 corresponding for instance to today. In the Einstein-De Sitter case, when $\Omega = 1$, the solution is $a \propto t^{2/3}$. The general solution is then a linear superposition of growing and decaying modes, $D_a$ and $D_b$,

$$\delta^{(1)} = k_a\, D_a + k_b\, D_b,$$

with

$$D_a = t^{2/3}, \quad \text{and} \quad D_b = t^{-1}. \tag{4}$$

The corresponding velocity field (actually its divergence) is deduced from the conservation equation (2), and the constants $k_a$ and $k_b$ are determined by the initial conditions for the density (or the gravitational potential) and the velocity fields.

For an open model with $\Omega < 1$, the parametric solution of Friedman equation is

$$a = A(\cosh\eta - 1), \quad \text{and} \quad t = B(\sinh\eta - 1),$$

with

$$A = \frac{4\pi G}{3}\overline{\rho}a^3\,|R|^2 = \frac{a_0}{2\left|1 - \Omega_0^{-1}\right|}, \qquad B = A\,|R|. \tag{5}$$

It is then convenient to use $y \equiv |\Omega^{-1} - 1|$ for time variable, in which case

$$D_a = 1 + \frac{3}{y} + \frac{3(1+y)^{1/2}}{y^{3/2}}\ln\left[(1+y)^{1/2} - y^{1/2}\right],$$

$$D_b = \frac{(1+y)^{1/2}}{y^{3/2}}. \tag{6}$$

Finally, the closed model case obtains by transforming the equations $(\eta, A, B, R, y) \to (i\eta, -A, iB, iR, -y)$.

If one now keeps also the $\mathcal{O}(\varepsilon^2)$ terms in eq. (3), one gets

$$\partial_{t^2}\delta + 2\frac{\partial_t a}{a}\delta = 4\pi G a^2 \overline{\rho}\delta(1+\delta) + \nabla\delta \cdot \frac{\nabla\phi}{a^2} + \frac{1}{a^2}\partial_\alpha\partial_\beta(v_\alpha v_\beta)$$

The solution in the Einstein–De Sitter case is found by using the first order solution (4)

$$\delta^{(2)} = \frac{5}{7}\delta^{(1)\,2} - \frac{1}{4\pi}\delta^{(1)}_{,\alpha}\Delta_{,\alpha} + \frac{1}{56\pi^2}\Delta_{,\alpha\beta}\Delta_{,\alpha\beta} \ ,$$

where $\Delta$ stands for $-\phi/(G\overline{\rho}a^2)$. As was noted by Martel and Freudling (1991), the equation (3) does not appear separable for $\Omega \neq 1$. But they found numerically a solution by assuming $\Omega^{1.2} \approx \Omega$. We shall in the following derive the exact analytical solution found by Bouchet et al. (1992) for arbitrary values of $\Omega$, in the absence of any cosmological constant. We shall also give an approximation of the third order solution, including the $\Lambda \neq 0$ case, in terms of usual mathematical functions.

*1.2. Developments of the Lagrangian Approach*

Zeldovich (1970) proposed to overcome some of the difficulties of the (pressureless) Eulerian approach by using the linear theory in terms of Lagrangian coordinates. The primary object of the analysis is then the trajectory of a "particle" instead of the density contrast. A fluid element (a particle) is indexed by its unperturbed Lagrangian coordinate $\boldsymbol{q}$. Its comoving Eulerian position at time $t$, $\boldsymbol{x}(t)$, is connected to $\boldsymbol{q}$ by a displacement field $\boldsymbol{\Psi}$

$$\boldsymbol{x} = \boldsymbol{q} + \boldsymbol{\Psi}(t, \boldsymbol{q}). \tag{7}$$

Zeldovich kinematic approximation amounts to assume that the displacement field is simply proportional to the initial displacement field, $\boldsymbol{\Psi}^i(\boldsymbol{q})$,

$$\boldsymbol{\Psi}(t, \boldsymbol{q}) = g_1(t)\boldsymbol{\Psi}^i(\boldsymbol{q}),$$

which is indeed a self-consistent irrotational solution of the linearized equations of motion, provided $g_1(t)$ is given by the linear growth rate (4). In effect, this is a ballistic approximation where gravitational acceleration is ignored; the movement is simply inertial.

The idea was to use this ansatz, even when the density contrast $\delta$ becomes large. It lead to the development of pancake theory. Indeed, the density constrast is given by the determinant, $J$, of the Jacobian of the transformation from $\boldsymbol{x}$ to $\boldsymbol{q}$,

$$\delta = \frac{1}{J} - 1 \ ,$$

with $J = |\det \mathcal{D}|$, $\mathcal{D}$ being the tensor of deformation,

$$\mathcal{D} = \left[\frac{\partial \boldsymbol{x}}{\partial \boldsymbol{q}}\right] = \mathcal{I} + \mathcal{T}, \quad \mathcal{T} = \left[\frac{\partial \boldsymbol{\Psi}}{\partial \boldsymbol{q}}\right], \tag{8}$$

where $\mathcal{I}$ is the identity matrix. If we call $\Lambda_i$ the eigenvalues of $\mathcal{T}/\}_\infty$, then $J = \prod_{i=1}^{3}(1+g_1(t)\Lambda_i)$, and there is a pancake forming perpendicular to the principal axis of $\mathcal{D}$ with the largest negative eigenvalue (if any). The spatial structure of perturbations near maxima of the density field (and the connection to angular momentum of galaxies and clusters) was thoroughly investigated by Doroshkevich (1970). Further developments along these lines are reviewed by Zeldovich (1978) and Doroshkevich, Shandarin, and Saar (1978).

Of course, as stated by Zeldovich, "the analytic evaluation of the error is extremely difficult". This lead to many comparisons with the exact dynamics, either with simulations (e.g., Doroshkevich et al. 1980) or rigorous Eulerian perturbative theory (Grinstein and Wise 1986). The approximation turns out to be amazingly good, at least when the initial field is smooth enough. Indeed, it is widely used today, e.g., to predict the weakly non-linear evolution of the moments of the one point probability distribution function of the density field (Betancort-Rijo 1991; see also Hoffman 1987 for the variance only), or of the distribution itself (Kofman et al. 1993, Padmanabhan & Subramanian 1993), or in reconstruction methods to recover the "initial conditions" from present day observations (e.g., Nusser et al. 1991, Nusser and Dekel 1992, Gramman 1992, Lachièze-Rey 1993a). In fact, it is even used rather often to address questions where it is not appropriate.

The success of Zeldovich approximation brought about many attempts to do better by correcting its shortcomings. A limitation of Zeldovich approximation is that, being linear (i.e., neglecting the gravitational acceleration induced by the changing density in the course of evolution), pancakes, once formed, indefinitely thicken in the absence of a restoring force (and also collapse times are overestimated). This can be vividly illustrated by examining Figure 1 which shows particle trajectories in a simple toy model. Following Moutarde et al. (1991) and Buchert (1993) this model is the superposition of 2 orthogonal sine waves with the same amplitude and wavelength, in a flat expanding Universe. This toy model is implemented in a square box, the axes of which are parallel to the waves and their comoving length, equal to the wavelength, is set to unity. $64^2$ particles are initially laid down on a regular lattice and at $a = 1$ their positions and velocities are perturbed, with an initial displacement field, $\boldsymbol{\Psi}^i$ given by

$$\begin{cases} \Psi^i_x(x,y) = \frac{\delta_i}{2\pi}\sin(2\pi x), \\ \Psi^i_y(x,y) = \frac{\delta_i}{2\pi}\sin(2\pi y), \quad \delta_i = \frac{1}{20}, \end{cases} \tag{9}$$

and $(d\boldsymbol{x}/da)_i = \boldsymbol{\Psi}^i$. The numerical simulation of the exact evolution was done with a PM code using a $128^2$ grid. With such initial conditions, particles move toward the center of the box, creating an overdensity and a potential well in which they will oscillate. The figure shows $x(a)$ for the particles initially closest to $y = 1/2$. According to Zeldovich approximation a particle with coordinate $\mathbf{q}$ before the perturbation has an Eulerian coordinate

$$\boldsymbol{x}(a, \boldsymbol{q}) = \boldsymbol{q} + a\boldsymbol{\Psi}^{\rm i}(\boldsymbol{q}) \qquad (10)$$

at expansion factor $a$. This results in straight line trajectories, with an overestimate of the collapse time, and artificial post-collapse thickening of the non-linear structures.

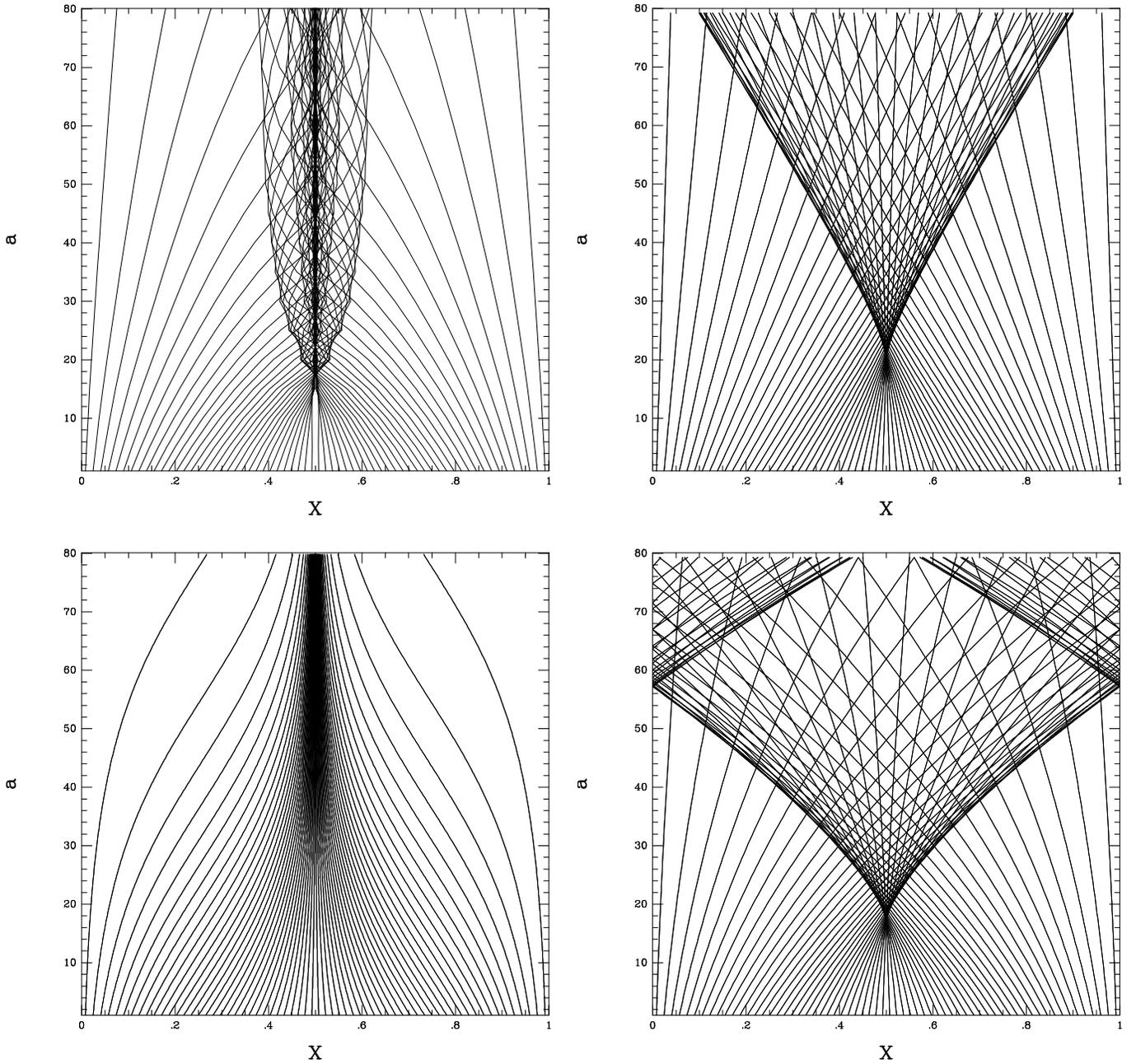

**Fig. 1.** Trajectories of particles for 2 in-phase, perpendicular, sine waves. The exact trajectories computed by direct numerical simulation are shown at the top left. The top right panel shoes the trajectories in Zeldovich approximation. The bottom panel shows the trajectories obtained by using the frozen flow approximation (left), and the second-order Lagrangian approximation (right).

A popular remedy to the second shortcoming of Zeldovich approximation is the "adhesion" approximation introduced by Gurbatov and Saïchev (1981, 1984) which forces particles orbits not to cross by introducing an artificial viscosity. This leads to Burger's equation of non-linear diffusion (Burgers 1940, 1974). It was later developed by many others, in particular Gurbatov, Saïchev, and Shandarin (1985, 1989), and Shandarin (1987); see the review by Shandarin and Zeldovich (1989) for further references. It has mainly been used to study the very large scale structures: for instance, Nusser and Dekel (1990) used this approach to study numerically their filamentary aspect in scale-invariant models, while Weinberg and Gunn (1990) focused on CDM initial conditions. For recent and more complete references, see Sahni et al. 1994 and Vergassola et al. 1994).

More recently, Matarese et al. (1992) suggested to improve Zeldovich approximation by replacing it with a "frozen flow" approximation. In Zeldovich solution, particles always keep the same velocities; these velocities are in effect "frozen". Matarese and its colleagues proposed instead to freeze the initial velocity *field*: particles move along streamlines computed (once and for all) from the initial potential; consequently, they never even reach the singularities. The figure 1 shows the corresponding trajectories in our toy model. Clearly, the pre-shell-crossing trajectories are not well described, but the approximation does capture some features of the asymptotically late evolution. A further improvement along this line of thought has just been proposed by Bagla and Padmanabhan (1994) who propose to freeze instead the initial potential (which is linearly conserved).

A much more ambitious program would be to try to find exact solutions to the full Lagrangian equations (Buchert and Götz 1987), maybe restricted to specific initial conditions. The only tractable case so far (Buchert 1989) is when the motion is locally one-dimensionnal, when two eigenvalues are (locally) zero. This is not too surprising, since Zeldovich approximate solution is well-known to be in fact exact for a purely one-dimensionnal collapse. More recently, Lachièze-Rey (1993) found a new class of formal solutions.

To our knowledge, the apparently obvious idea of using a rigorous Lagrangian perturbative expansion is due to Moutarde et al. (1991). These authors, in the course of a systematic study of smooth and isolated perturbations, first found numerically that the evolution of three sine waves leads, at shell crossing (when orbits intersect), to a scale-invariant density profile; their results were thus extending the two-dimensionnal result of Alimi et al. (1990). They reasoned that this could be analytically confirmed by a Lagrangian approach, in the same spirit that lead Zeldovich to his approximation, since this power-law density profile builds up before shell crossing. They performed a Lagrangian perturbative expansion up to the third order, in the $\Omega = 1$ case. Their determination of the spatial part of the solution, and of the growth rates up to the third order, was only performed explicitly for the initial conditions of their simulations. They found exact agreement with the numerical density profile near the collapse time up to a density contrast of order 50! Figure 1 shows the trajectories for our toy model with sine waves as computed with the second-order lagrangian approximation (hereafter referred to as L2). Indeed, as long as particle trajectories do not cross, even at very late stages, their trajectories are correctly described by the second order approximation. Consider for instance the particles initially close to $x = 0$, their position up to $a \approx 80$ are well predicted. At the same time, trajectories of particles initially closer to $x = 1/2$ remain correct up to the time of collapse and therefore the time of first shell crossing is more correctly estimated by L2 than by Zeldovich approximation. On the other hand, Zeldovich approximation gives better results after shell crossing because particles go apart from the point of collapse more slowly than in L2. One can see that at $a = 80$ structures are completely washed out in the L2 approximation. However, one can estimate that, as far as $a \approx 30$, a bit after shell crossing, L2 is closer to the simulation than Zeldovich. Thus figure 1 clearly shows what to expect for the various approximations: the frozen flow captures some features of the asymptotically late behavior, while L2 improves the description of the weakly non-linear phase.

The work of Moutarde et al. (1991) was soon generalized in two ways. On one hand, Buchert (1992) explored the perturbative solutions for arbitrary initial conditions (in the $\Omega = 1$ case). He found for instance that Zeldovich solution is a sub-class of the general first order solutions, when peculiar velocities and accelerations are required to be parallel (which is true for irrotational flows at late times, when the growing mode dominates, as was explicitly assumed by Zeldovich from start). He also explored the first order solutions for rotational perturbations.

On the other hand, Bouchet et al. (1992) explored a more restricted class (the same class of solution as Zeldovich's, namely the fastest growing part of an irrotational flow), but gave the second-order exact solution for Friedman-Lemaître models of arbitrary density parameter $\Omega$, provided there is no cosmological constant (the first order, i.e., Zeldovich solution for this case has been known for long, as in the Eulerian case, see Shandarin 1980; one even knows the behavior with a cosmological constant, Bildhauer et al. 1992). And as an illustration of the power of the approach, Bouchet et al. (1992) made a first connection with measurable statistical quantities, by computing the value of the skewness of the density field probability distribution function (PDF). For $\Omega = 1$, it is a constant, $S_3 = 34/7$, times the square of the variance of the PDF, as was already found by Peebles (1980). They showed that $S_3$ depends extremely weakly on $\Omega$ (see below). And that $S_3$ is barely affected by a real space-redshift space mapping. This is brought further by Juszkiewicz et al. (1993; see also Juszkiewicz and Bouchet 1991) who consider the effect of smoothing, and by a companion paper by Hivon et al. (1994) who consider the effect of smoothing in redshift space. A recent summary of most of these results may be found in Bouchet and Juszkiewicz (1993).

Quite recently, Buchert and Ehlers (1993) and Buchert (1993) explore (respectively) the generic second and third order solutions in the $\Omega = 1, \Lambda = 0$ case, while Lachièze-Rey (1993) propose a tensorial reformulation of the equations of Bouchet et al. (1992) which naturally includes the curl-free and divergence-free part of the vector field. Interestingly, Lachièze-Rey found a special tensorial solution which is purely local. Unfortunately, it is not a general solution. Indeed, an alternative view of the problem is given by its relativistic generalization which was reviewed by Ellis (1971). It was then found that a quantity called the magnetic part of the Weyl tensor vanishes in the Newtonian limit (when keeping the highest order term in a $1/c$ expansion). The equations are then quite simple and, assuming this holds at higher orders, Barnes and Rowlingson (1989), Matarese, Pantano, and Saez (1993), Bertschinger (1993), Croudace et al. (1994) obtained closed sets of *local* Lagrangian perturbative equations.

This generated a lot of excitement, as it seemed to magically evade the non-local character of the perturbative Newtonian approach. Bertschinger and Jain (1994) proceeded to deduce that, contrary to Zeldovich pancaking picture, *typical* perturbations would rather collapse as filaments. Quite recently, though, it has been realized that this is incorrect, since at the post-newtonian level (i.e. at higher orders of a $1/c$ expansion) the magnetic part of the Weyl tensor is non-vanishing and should be included in a non-linear treatment of the post-Newtonian limit (Kofman and Pogosyan 1994; see also Bertschinger and Hamilton 1994). Indeed, a correct $1/c$ expansion of the equations of General Relativity does lead to the same equations as in the standard perturbative Newtonian theory, implying that the Lagrangian evolution is *not* purely local.

## 2. Lagrangian Perturbation Theory

The Lagrangian perturbative approach introduced by Moutarde et al. (1991) proceeds in quite a parallel fashion to the Eulerian case recalled in the introduction. As before, we use the Newtonian approximation with comoving coordinates, but following Doroshkevich et al. (1973) we replace the standard cosmological time by a new time $\tau$ defined by

$$d\tau \propto a^{-2} dt. \tag{11}$$

In that case the motion and field equations now read

$$\ddot{\boldsymbol{x}} = -\nabla_x \Phi, \quad \Delta_x \Phi = \beta(\tau)\delta, \tag{12}$$

with no $\dot{\boldsymbol{x}}$ term. In this equation, dots denote derivatives with respect to $\tau$, and $\beta(\tau) = 4\pi G a (\overline{\rho} a^3)$. By choosing the proportionality constant in the definition (11) to be $-1, A/|R|$, or $-A/|R|$ when, respectively, $\Omega$ is equal, smaller or greater than 1, $\beta$ is simply given by

$$\beta = \frac{6}{\tau^2 + k(\Omega)},$$

with

$k(\Omega = 1) = 0$
$k(\Omega < 1) = -1$
$k(\Omega > 1) = +1.$

While $\tau = t^{-1/3}$, when $\Omega = 1$, one has $\tau = |1 - \Omega|^{-1/2}$ otherwise.

Now we wish to follow the particle trajectories instead of the density contrast, by using the mapping (7). The Jacobian of the transformation from $\boldsymbol{x}$ to $\boldsymbol{q}$, permits to express the requirement of mass conservation simply as

$$\rho(\boldsymbol{x}) J d^3q = \rho(\boldsymbol{q}) d^3 q,$$

i.e., $\delta = J^{-1} - 1$. By taking the divergence of the equation of motion (12), one obtains the equivalent of the Eulerian equation (3):

$$J(\tau,\boldsymbol{q})\nabla_x \ddot{\boldsymbol{x}} = \beta(\tau)\left[J(\tau,\boldsymbol{q}) - 1\right]. \tag{13}$$

Of course the addition of any divergence-free displacement field to a solution of the previous equation will also be a solution. In the following, we remove this indeterminacy by restricting our attention to potential movements, which must satisfy

$$\nabla_x \times \ddot{\boldsymbol{x}} = 0. \tag{14}$$

The main reason to restrict to that case is that vortical perturbations linearly decay, a consequence of the conservation of angular momentum in an expanding universe. Thus one might consider that the solutions will apply anyway, even if vorticity is initially present, because at later times it will decay away. In the same spirit, we shall mainly focus on growing mode solution (see Buchert and Ehlers (1993) and Buchert (1993) for the cases of rotational perturbations and the effect of decaying modes).

The final equation to solve obtains by rewriting the divergence of the acceleration $\Gamma \equiv \ddot{\boldsymbol{x}}$ explicitly as a function of $\boldsymbol{q}$

$$\nabla_x \Gamma = J(\tau,\boldsymbol{q})^{-1} \sum_{i,j} \Gamma_{i,j} A_{ji}, \tag{15}$$

where the $A_{ij}$ are the cofactors of the Jacobian, and the partial derivatives denoted by latin letter are taken with respect to the $\boldsymbol{q}$ coordinate (e.g., $\Gamma_{i,j} = \frac{\partial \Gamma_i}{\partial q_j}$).

As in the Eulerian case, perturbative solutions are obtained by means of an iterative procedure. But this time, the expansion concerns the particles displacement field itself, and we write it as

$$\boldsymbol{\Psi} = \varepsilon \boldsymbol{\Psi}^{(1)} + \varepsilon^2 \boldsymbol{\Psi}^{(2)} + \varepsilon^3 \boldsymbol{\Psi}^{(3)} + \mathcal{O}(\varepsilon^4). \tag{16}$$

The determinant of the jacobian is then similarly expanded as

$$\begin{aligned} J &= 1 + \varepsilon J^{(1)} + \varepsilon^2 J^{(2)} + \varepsilon^3 J^{(3)} + \mathcal{O}(\varepsilon^4) \\ &= 1 + \varepsilon K^{(1)} + \varepsilon^2 (K^{(2)} + L^{(2)}) \\ &\quad + \varepsilon^3 (K^{(3)} + L^{(3)} + M^{(3)}) + \mathcal{O}(\varepsilon^4) \end{aligned} \tag{17}$$

where $K^{(m)}$, $L^{(m)}$, and $M^{(m)}$ denote the $m-th$ order part of the (invariant) scalars

$$\begin{cases} K &= \nabla \cdot \boldsymbol{\Psi} = \sum_i \Psi_{i,i} \\ L &= \frac{1}{2} \sum_{i \neq j} (\Psi_{i,i} \Psi_{j,j} - \Psi_{i,j} \Psi_{j,i}) \\ M &= \mathcal{D} = \det[\Psi_{i,j}] \ . \end{cases} \quad (18)$$

In other words,

$$\begin{cases} K^{(m)} &= \nabla \cdot \boldsymbol{\Psi}^{(m)} = \sum_i \Psi_{i,i}^{(m)} \\ L^{(2)} &= \frac{1}{2} \sum_{i \neq j} (\Psi_{i,i}^{(1)} \Psi_{j,j}^{(1)} - \Psi_{i,j}^{(1)} \Psi_{j,i}^{(1)}) \\ L^{(3)} &= \sum_{i \neq j} (\Psi_{i,i}^{(2)} \Psi_{j,j}^{(1)} - \Psi_{i,j}^{(2)} \Psi_{j,i}^{(1)}) \\ M^{(3)} &= \det[\Psi_{i,j}^{(1)}] \end{cases} \quad (19)$$

The equation to solve then follows by replacing the expansions (16-18) of the displacement field and of the Jacobian in equations (13) and (15). We now proceed to an iterative solution of this equation.

### 2.1. First Order

First, keeping only the terms of order $\varepsilon$, one gets

$$\ddot{K}^{(1)} - \beta K^{(1)} = 0. \quad (20)$$

Thus $K^{(1)}(\tau, \boldsymbol{q})$ can be factorised into a temporal and spatial part,

$$K^{(1)}(\tau, \boldsymbol{q}) = g_1(\tau) K^{(1)}(\tau_i, \boldsymbol{q}),$$

where $K^{(1)}(\tau_i, \boldsymbol{q})$ is the divergence of the initial displacement field, $g_1(\tau_i)$ is assumed to be unity, and

$$g_1 = k_a g_{1a} + k_b g_{1b}. \quad (21)$$

For $\Omega = 1$, the linear growth rate $g_1$ is given by

$$g_{1a} = \tau^{-2}, \quad g_{1b} = \tau^3. \quad (22)$$

When $\Omega < 1$,

$$g_{1a} = 1 + 3(\tau^2 - 1)(1 + \tau S), \quad g_{1b} = \tau(\tau^2 - 1), \quad (23)$$

$$S = \frac{1}{2} \ln \frac{\tau - 1}{\tau + 1} \ . \quad (24)$$

The closed case of course obtains by the transformation $(\eta, A, B, R, x, \tau, \beta) \to (i\eta, -A, iB, iR, -x, i\tau, -\beta)$.

For a potential movement, one thus recovers Zeldovich solution (1970)

$$\boldsymbol{\Psi}^{(1)}(\tau, \boldsymbol{q}) = g_1(\tau) \tilde{\boldsymbol{\Psi}}^{(1)}(\boldsymbol{q}), \quad (25)$$

if we define (for all $m$)

$$\tilde{\boldsymbol{\Psi}}^{(m)}(\boldsymbol{q}) \equiv \boldsymbol{\Psi}^{(m)}(\tau_i, \boldsymbol{q}). \quad (26)$$

Also, note that

$$\delta^{(1)}(\tau) \propto g_1(\tau)$$

(with initially $\delta_i^{(1)} = -\nabla \cdot \tilde{\boldsymbol{\Psi}}^{(1)}(\boldsymbol{q})$), i.e., the Eulerian linear behavior is recovered, since $g_1 \equiv D$ [cf. Eqs. (4), (6)].

The logarithmic derivative of the growth factor,

$$f_1 \equiv \frac{a}{g_1} \frac{dg_1}{da} \ , \quad (27)$$

is useful to describe comoving peculiar velocities and therefore redshift distortion (see below). Near $\Omega = 1$, a limited expansion of the solution (24) shows that $f_1 \asymp \Omega^{4/7}$ where $\asymp$ means "behaves asymptotically as". A better analytical fit for $f_1$ in the range $0.1 < \Omega < 1$ is given by $f_1 \approx \Omega^{3/5}$, as was originally proposed by Peebles (1976).

### 2.2. Second Order

Collecting terms of $\mathcal{O}(\varepsilon^2)$ in equations (13) and (15), we find

$$\ddot{K}^{(2)} - \beta K^{(2)} = -\beta g_1^2 L^{(2)}(\tau_i, \boldsymbol{q}). \quad (28)$$

Thus, the solution is again separable

$$K^{(2)}(\tau, \boldsymbol{q}) \equiv g_2(\tau) K^{(2)}(\tau_i, \boldsymbol{q}) \quad (29)$$

where $g_2(\tau_i) = 1$. The spatial part is given by

$$K^{(2)}(\tau_i, \boldsymbol{q}) = L^{(2)}(\tau_i, \boldsymbol{q}),$$

and the growth factor is

$$g_2 = k_a^2 g_{2a} + 2 k_a k_b g_{2b} + k_b^2 g_{2c} + l_a g_{1a} + l_b g_{1b} \ . \quad (30)$$

For $\Omega = 1$, we find

$$g_{2a} = -\frac{3}{7} g_{1a}^2, \quad g_{2b} = \frac{3}{2} g_{1a} g_{1b}, \quad g_{2c}^2 = -\frac{1}{4} g_{1b}. \quad (31)$$

When $\Omega < 1$,

$$\begin{aligned} g_{2a} &= 1 - \frac{9}{4}(\tau^2 - 1)\left[\tau + (\tau^2 - 1)S\right]^2 , \\ g_{2b} &= -\frac{3}{4}(\tau^2 - 1)\left[\tau^3 + (\tau^2 - 1)S\right] , \quad (32) \\ g_{2c} &= -\frac{1}{4}(\tau^2 - 1)^3 . \end{aligned}$$

The closed case, once again, obtains by the transformation $(\eta, A, B, R, x, \tau, \beta) \to (i\eta, -A, iB, iR, -x, i\tau, -\beta)$.

If only a growing mode is initially present ($k_b = 0 = l_b$), which we assume, one must have $g_2/g_1 \to 0$ when $\tau \to \infty$, i.e., when $\Omega \to 1$. A limited expansion of equation (32) then shows that $l_a$ must be equal to $-\frac{3}{2}$, and this yields the physically relevant solution

$$g_2 = -\frac{1}{2} - \frac{9}{2}(\tau^2 - 1)\left\{1 + \tau S + \frac{1}{2}\left[\tau + (\tau^2 - 1)S\right]^2\right\}, \quad (33)$$

which behaves near $\Omega = 1$ as

$$g_2 \asymp -\frac{3}{7} \Omega^{-2/63} g_1^2, \quad (34)$$

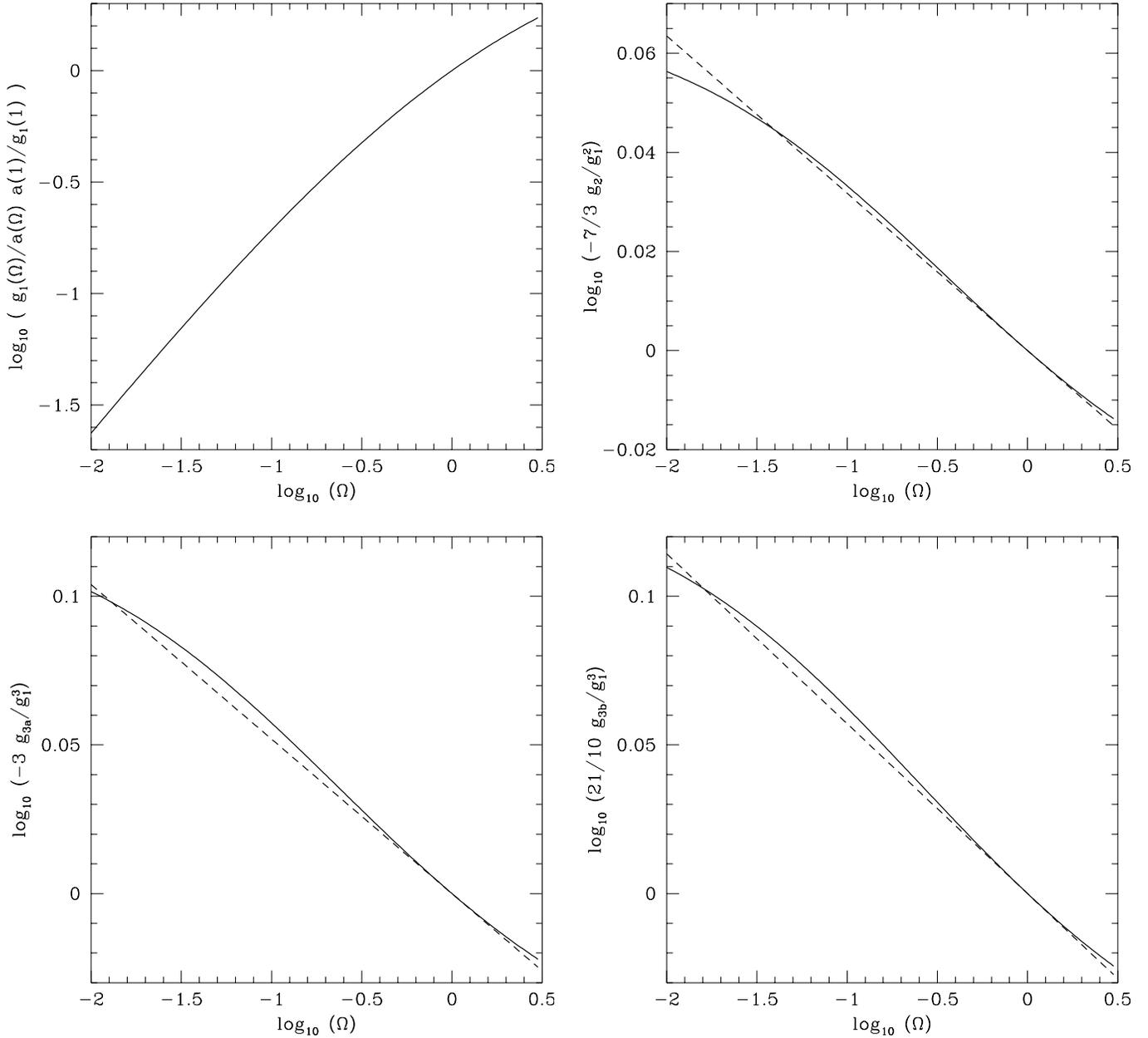

**Fig. 2.** Growth factors for Friedman-Lemaître models with arbitrary $\Omega$ and $\Lambda = 0$. The upper panels display the first order growth rate divided by the expansion factor (left), and the second order one (right), while the lower panels show the two third order growth factors. The solid curves show the exact solutions and the dashed curves correspond to the asymptotic behaviors near $\Omega = 1$.

while $g_2/g_1^2 \to -1/2$ when $\Omega$ approaches 0.

If $\Omega > 1$, the solutions are obtained by the usual transformation, and in particular, the solution corresponding to (33) is

$$g_2 = -\frac{1}{2} + \frac{9}{2}(\tau^2 + 1)\left\{1 - \tau S - \frac{1}{2}\left[\tau - (\tau^2 + 1)S\right]^2\right\},$$
$$S = \text{Arctg}\frac{1}{\tau}, \tag{35}$$

with $g_2$ approaching $4 - (3\pi/4)^2$ (while $g_1 \asymp 2$) when $\Omega$ goes to infinity (i.e., $\tau \to 0$), and with the same behavior (34) near $\Omega = 1$.

Figure 2 shows $-\frac{7}{3}g_2/g_1^2$ versus $\Omega$, and it confirms that indeed $g_2/g_1^2$ is very nearly constant, as could be foreseen from the asymptotical behaviors. It is thus not surprising that the remaining small variation of $g_2$ may be described quite accurately by Eq. (34) for $0.1 \lesssim \Omega \lesssim 2$, which should cover most astrophysical uses (Bouchet et al. 1992).

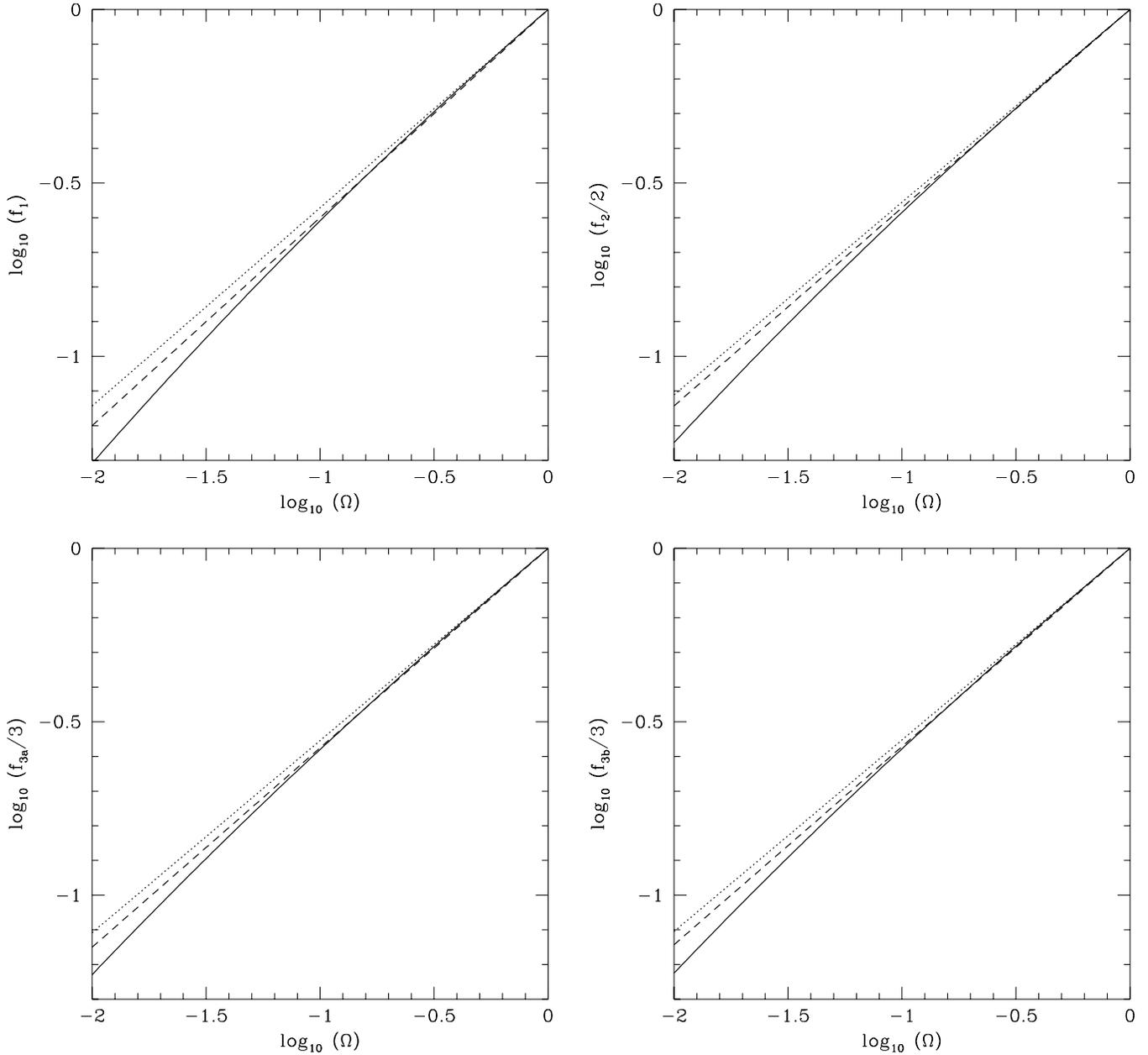

**Fig. 3.** Logarithmic derivatives of the growth factors of Fig. 2, i.e. for Friedman-Lemaître models with arbitrary $\Omega$ and $\Lambda = 0$. The upper panels correspond to the first order (left) and second order (right). The lower panels correspond to the two third order factors. The solid curves show the exact solutions, the dotted lines display their slope in $\Omega = 1$, and the dashed ones a better fit for $0.1 \lesssim \Omega \lesssim 1$.

The logarithmic derivative of the second order growth rate,

$$f_2 \equiv \frac{a}{g_2} \frac{dg_2}{da} \,,$$

behaves as $f_2 \asymp 2\,\Omega^{5/9}$ near $\Omega = 1$. Alternatively, it can be somewhat better approximated by $2\,\Omega^{4/7}$ for $0.1 \lesssim \Omega \lesssim 1$ (see Fig. 3).

### 2.3. Third Order

The $\mathcal{O}(\varepsilon^3)$ terms obey the equation

$$\ddot{K}^{(3)} - \beta K^{(3)} = -2\beta g_1^3 M^{(3)}(\tau_i) + \beta g_1^3 (1 - \frac{g_2}{g_1^2}) L^{(3)}(\tau_i), \quad (36)$$

which is separable only when $\Omega = 1$. In that case, the fastest growing solution is simply

$$K^{(3)}(\tau, \boldsymbol{q}) = -\frac{1}{3}\, g_1^3(\tau) \left[ M^{(3)}(\tau_i, \boldsymbol{q}) - \frac{5}{7}\, L^{(3)}(\tau_i, \boldsymbol{q}) \right] \,,$$

Otherwise the third order term may be written as:

$$K^{(3)}(\tau, \boldsymbol{q}) = g_{3a}(\tau) M^{(3)}(\tau_i, \boldsymbol{q}) + g_{3b}(\tau) L^{(3)}(\tau_i, \boldsymbol{q}), \quad (37)$$

assuming as before that $g_{3a}(\tau_i) = g_{3b}(\tau_i) = 1$. Figure 2 shows these growth rates, $g_{3a}$ and $g_{3b}$, which we obtained by numerical integration of equation (36). Alternatively, by expanding the equations governing the ratios $g_{3a}/g_1^3$ and $g_{3b}/g_1^3$, we find that the fastest growing parts behave near $\Omega = 1$ as

$$g_{3a} \asymp -\frac{1}{3} \Omega^{-4/77} g_1^3, \quad g_{3b} \asymp \frac{5}{21} \Omega^{-2/35} g_1^3 . \quad (38)$$

Fig. 2 shows that these asymptotic behaviors near $\Omega = 1$ provide very good fits to $g_3/g_1^3$, as was also the case before for the ratio $g_2/g_1^2$ [Eq. (34)]. In both cases the $\Omega$ dependence of the ratios is very weak.

The logarithmic derivatives of the growth factors $f_i = (a/g_i) d g_i/da$ behave near $\Omega = 1$ as

$$f_{3a} \asymp 3\,\Omega^{128/231}, \quad f_{3b} \asymp 3\,\Omega^{58/105},$$

or can be better approximated for $0.1 \lesssim \Omega \lesssim 1$ by $3\,\Omega^{23/40}$ and $3\,\Omega^{4/7}$ respectively (see Fig. 3).

### 2.4. Non-zero Lambda

We now turn to an inflation motivated case, i.e. a flat Universe with a non-zero cosmological constant,

$$\Omega + \frac{\Lambda}{3H^2} = 1, \quad (39)$$

The expansion factor and the Hubble parameter write

$$a(t) = a_1 \sinh^{2/3}\left(\frac{3}{2} H_\infty t\right), \quad (40)$$

$$H(t) = H_\infty \coth\left(\frac{3}{2} H_\infty t\right), \quad (41)$$

where $a_1$ is the expansion factor such that $\Omega = 1/2$ and $H_\infty \equiv \sqrt{\Lambda/3} = H(t = \infty)$. It is convenient to introduce the inflexion point $a_e = 2^{-1/3} a_1$. For $a$ smaller than $a_e$ the model-universe is dominated by matter with a negative acceleration, whereas for larger $a$, it's evolution is dominated by the $\Lambda$ term, the acceleration is positive and continuously increases. Introducing the new variable $h = H(t)/H_\infty$, the equation of motion for a particle with comoving coordinates $\boldsymbol{x}$ now reads

$$\nabla_x \left( 3(h^2 - 1) \frac{d^2}{dh^2} + 2h \frac{d}{dh} \right) \boldsymbol{x} = -2\delta, \quad (42)$$

with $d\boldsymbol{x}/dh \equiv \partial \boldsymbol{x}/\partial h|_q$. Applying the same techniques as in the previous part, we find the equations satisfied by the growth rates up to the third order

$$\begin{cases} 3(h^2-1)\ddot{g}_1 + 2h\dot{g}_1 = 2g_1 \\ 3(h^2-1)\ddot{g}_2 + 2h\dot{g}_2 = 2g_2 - 2g_{1+}^2 \\ 3(h^2-1)\ddot{g}_{3a} + 2h\dot{g}_{3a} = 2g_{3a} - 2g_{1+}^3 \\ 3(h^2-1)\ddot{g}_{3b} + 2h\dot{g}_{3b} = 2g_{3b} + 2g_{1+}^3(1 - \frac{g_{2+}}{g_{1+}^2}) \end{cases} \quad (43)$$

where dots denote derivatives with respect to $h$. Of course, the spatial parts remain given by Eqs. (29), (37).

Solutions of the first order equation are linear combinations of the respectively growing and decreasing modes

$$g_{1+}(h) = h \int_h^\infty \frac{dh'}{h'^2 (h'^2 - 1)^{1/3}}, \quad \text{and} \quad g_{1-}(h) = h. \quad (44)$$

By expressing $h$ as a function of $x = a/a_e$, one can verify that these solutions are respectively identical with $D_1$ and $D_2$ given in LSS (eq. 13.6). The growing mode solution is an ugly mixture of hypergeometric functions, but we can identify its asymptotic behaviors:

$$g_{1+}(x) \propto x, \quad x \ll 1; \quad (45)$$

$$g_{1+}(x) \to \int_1^\infty \frac{du}{u^2 (u^2 - 1)^{1/3}}, \quad x \to \infty. \quad (46)$$

We have solved the differential equations (43) recursively by numerical integration, to find the fastest growing solutions whose behaviors are shown in figure 4.

Note that near $\Omega = 1$ (i.e., $h \to \infty$) the second order solution behaves as

$$g_2(\Omega) \asymp -3/7\,\Omega^{-1/143} g_1^2 .$$

Similarly, the two third order growth factors behave near $\Omega = 1$ as

$$g_{3a} \asymp -\frac{1}{3} \Omega^{-4/275} g_1^3, \quad g_{3b} \asymp \frac{10}{21} \Omega^{-269/17875} g_1^3 . \quad (47)$$

Figure 4 shows that these asymptotic solutions near $\Omega = 1$ yield good approximations for a large range of values of $\Omega$, $0.1 \lesssim \Omega \lesssim 1$.

The logarithmic derivatives of the growth factors ($f_i = (a/g_i) dg_i/da$) behave near $\Omega = 1$ as

$$f_1 \asymp \Omega^{6/11}, \quad f_2 \asymp 2\,\Omega^{153/286}, \quad (48)$$

$$f_{3a} \asymp 3\,\Omega^{146/275}, \quad f_{3b} \asymp 3\,\Omega^{9481/17875}. \quad (49)$$

However, a better analytical fit for $0.1 \lesssim \Omega \lesssim 1$ would be (see Fig. 5)

$$f_1 \approx \Omega^{5/9}, \quad f_2 \approx 2\,\Omega^{6/11}, \quad (50)$$

$$f_{3a} \approx 3\,\Omega^{13/24}, \quad f_{3b} \approx 3\,\Omega^{13/24} . \quad (51)$$

Values of $f_1$ for non vanishing cosmological constant have already been calculated for a flat Universe (Peebles 1984) or in the general case (Lahav et al. 1991, Martel 1991). Lahav et al. (1991) proposed the fit

$$f_1 \approx \Omega^{0.6} + \lambda/70(1 + \Omega/2) \quad (52)$$

with $\lambda = \Lambda/(3H^2)$, whereas Martel (1991) proposed the simpler fit

$$f_1 \approx \Omega^{0.6} + \lambda/30. \quad (53)$$

The computations and fits proposed by these authors are compared to our results in Fig. 5.

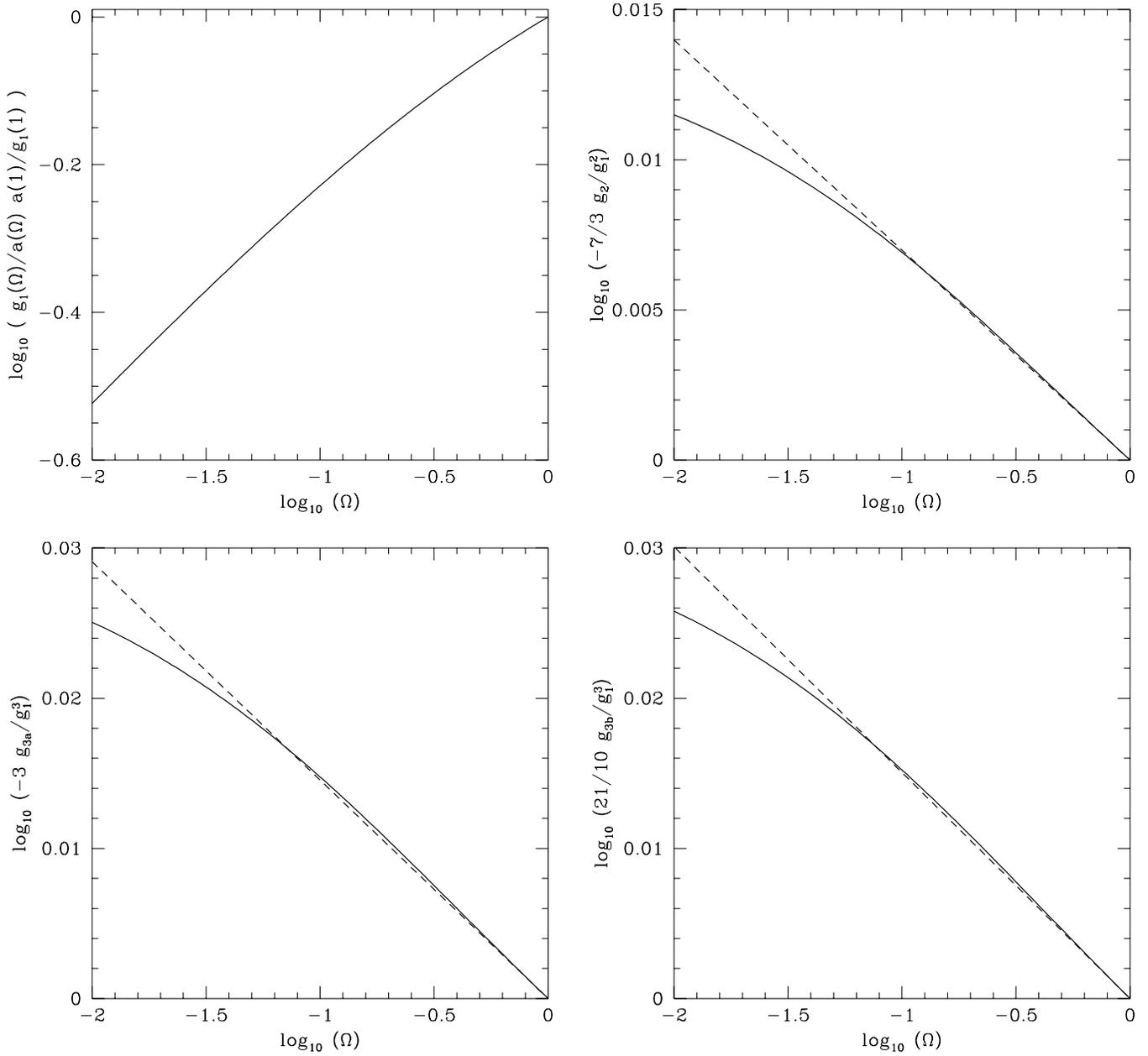

**Fig. 4.** Same as Fig. 2, but for a flat Friedman-Lemaître models with $\Lambda \neq 0$ (the upper left panel shows the first order growth divided by the expansion factor, the upper right one shows the second order growth rate, and the lower panels are devoted to the two third order growth factors. The solid curves show the exact solutions, and the dashed lines display the asymptotic behavior in $\Omega = 1$.

## 3. Link to statistics

In order to make the connection with measurable quantities, we proceed as follows. Let $Q(\boldsymbol{x})$ be an Eulerian physical quantity of interest. As usual, we shall assume that ensemble averages, denoted by $\langle Q \rangle$ are equivalent to averages over space, denoted by $\overline{Q}$,

$$\overline{Q}(x) = \frac{1}{V} \int_V d^3x \, Q(x),$$

if the volume $V$ is large enough that it can be considered a fair sample. Then

$$\langle Q(x) \rangle_{\mathrm{x}} \equiv \langle Q(\boldsymbol{q}) \, J(\boldsymbol{q}) \rangle_{\mathrm{q}}, \tag{54}$$

where the subscripts x or q tell us whether the average is to be taken in Eulerian or Lagrangian space.

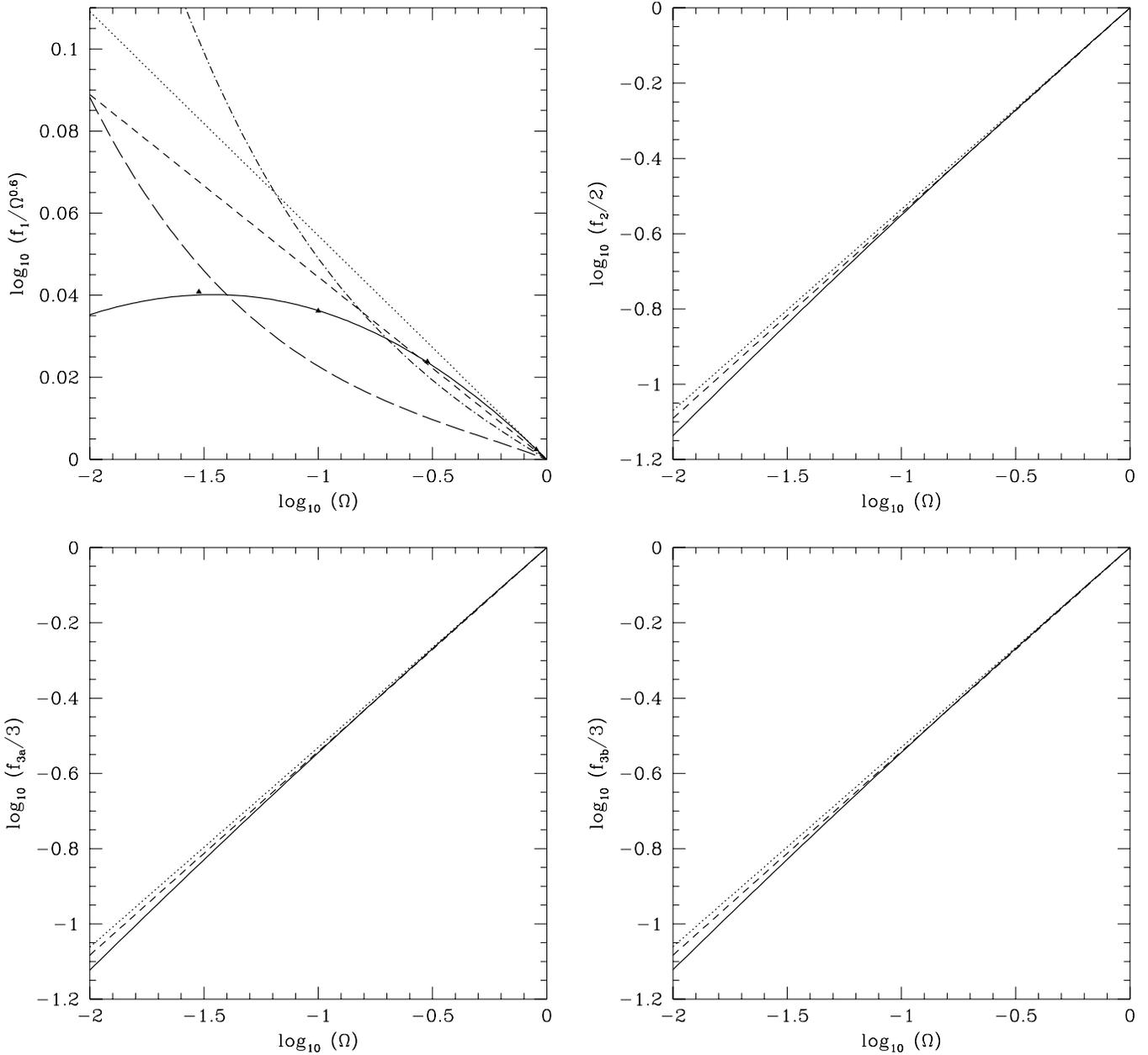

**Fig. 5.** Logarithmic derivatives of the growth factors for a flat Universe with $\Lambda \neq 0$. The upper panels correspond to the first order (left) and second order (right) rates. The lower panels correspond to the two third order factors. The solid curves show the exact solutions, the dotted lines display their slope in $\Omega = 1$ [see Eq. (49)], and the dashed ones a better fit for $0.1 < \Omega < 1$ [see Eq. (51)]. We also show for comparison in the upper left plot the values for $f_1$ provided by Peebles (triangles), Lahav et al. (long dashes), and Martel (dashes-dots).

## 3.1. Skewness factor in real space

Now that large galaxy samples are becoming available, it becomes of particular interest to predict the moments of the density contrast distribution

$$\langle \delta^n(\mathbf{x}) \rangle_{\mathrm{x}} = \left\langle (J^{-1} - 1)^n \right\rangle_{\mathrm{x}} = \left\langle (J^{-1} - 1)^n J \right\rangle_{\mathrm{q}}.$$

The variance and skewness are then given up to the fourth order by

$$\langle \delta^2 \rangle = \varepsilon^2 \left\langle J^{(1)\,2} \right\rangle + \varepsilon^3 \left\langle 2 J^{(1)} J^{(2)} - J^{(1)\,3} \right\rangle +$$
$$\varepsilon^4 \left\langle J^{(1)\,4} - 3 J^{(1)\,2} J^{(2)} + J^{(2)\,2} + 2 J^{(1)} J^{(3)} \right\rangle + \mathcal{O}(\varepsilon^5)$$
$$\langle \delta^3 \rangle = -\varepsilon^3 \left\langle J^{(1)\,3} \right\rangle + \varepsilon^4 \left\langle 2 J^{(1)\,4} - 3 J^{(1)\,2} J^{(2)} \right\rangle + \mathcal{O}(\varepsilon^5),$$

where all averages on the displacement field are taken with respect to the Lagrangian unperturbed coordinate q.

We consider the case of an initially non Gaussian density field $\delta_i = \varepsilon \delta^{(1)}$, but let us require for simplicity that the three components of the displacement field $\tilde{\boldsymbol{\Psi}}^{(1)}$ are independent, with the same statistical law to insure homogeneity and isotropy. We note $\sigma^2$ the variance of any component $i$ of the (linear) gradient field

$$\sigma^2 = \left\langle \Psi_{i,i}^{(1)\,2} \right\rangle = \varepsilon^2 \, g_1^2 \, \langle \delta_i^2 \rangle / 3,$$

and $S$ its third moment $S = \left\langle \Psi_{i,i}^{(1)\,3} \right\rangle$, and $K$ its reduced forth moment $K = \left\langle \Psi_{i,i}^{(1)\,4} \right\rangle - 3\sigma^4$. We thus have $\left\langle J^{(1)\,3} \right\rangle = 3S$, $\left\langle J^{(1)\,4} \right\rangle = 3K + 27\sigma^4 = 3K + 3\left\langle J^{(1)\,2} \right\rangle^2$. The other term in $\langle \delta^3 \rangle$ involves the product $J^{(1)\,2} J^{(2)}$ which can readily be estimated since [after (18), (29) and (19)] we have

$$J^{(2)} = (1 + g_2/g_1^2) \sum_{i>j} (\Psi_{i,i}^{(1)} \Psi_{j,j}^{(1)} - \Psi_{i,j}^{(1)} \Psi_{j,i}^{(1)}).$$

It follows by development that $\left\langle J^{(1)\,2} J^{(2)} \right\rangle = 6(1 + g_2/g_1^2)\sigma^4$. We thus have

$$S_3 = \frac{-3S + \varepsilon g_1 6(K + 3(2 - g_2/g_1^2)\sigma^4)}{\varepsilon g_1 \sigma^4 (3 - 3\varepsilon S/\sigma^2)^2}. \tag{55}$$

$$= -\frac{S}{3\sigma^4 g_1 \varepsilon} + 4 - 2\, g_2/g_1^2 + 2\, \frac{K\sigma^2 - S^2}{3\sigma^6} + \mathcal{O}(\varepsilon). \tag{56}$$

For an initially gaussian field with $S = K = 0$, we get the simple result

$$S_3 = 4 - 2\, g_2/g_1^2 + \mathcal{O}(\varepsilon^2), \tag{57}$$

whose first term corresponds to the pure Zel'dovich approximation and had been found by Grinstein and Wise (1987). The value of the ratio of growing modes may be obtained from the exact solutions given before (or read off Fig. 2), but it is hard to imagine practical cases when our approximation (34) might not be sufficient. We thus have the handy and quite accurate formula for a gaussian initial field with $0.1 \lesssim \Omega \lesssim 2$ (Bouchet et al. 1992)

$$S_3 \approx \frac{28 + 6\,\Omega^{-2/63}}{7},$$

which generalizes the $S_3 = 34/7$ found by Peebles (1976) in the $\Omega = 1$ case. The more general formula (56) gives the time-evolution of the skewness factor $S_3$ in spatially flat models for the class of non-gaussian initial conditions which may be generated by independent displacement fields along three axes. The general formula for the $\Omega = 1$ case can be found in Fry and Scherrer (1993).

### 3.2. Skewness factor in redshift space

We now turn to a generalization of the previous result (56) in redshift space. In redshift space, the appearance of structures is distorted by peculiar velocities. At "small" scales, this leads to the "finger of god" effect: the clusters are elongated along the line-of-sight due to their internal velocity dispersion. This is an intrinsically non-linear effect, and we shall not be concerned with it. At "large" scales, the effect is reversed: the coherent inflow leads to a density contrast increase parallel to the line-of sight. Indeed, foreground galaxies appear further than they are, while those in the back look closer, both being apparently closer to the accreting structure (Sargent & Turner 1977; LSS, §76; Kaiser 1987).

Kaiser (1987) estimated this redshift effect, in the large sample limit, on the direction averaged correlation function, and found $\left\langle \delta^2 \right\rangle^z = (1 + 2/3\,\Omega^{0.6} + 1/5\,\Omega^{1.2})\left\langle \delta^2 \right\rangle$, where the superscript $z$ correspond to a redshift space measurements. For $\Omega = 1$, $\left\langle \delta^2 \right\rangle^z = 28/15\,\left\langle \delta^2 \right\rangle$

Let us thus consider the case of spherical coordinates, when distances to the observer would be estimated by means of redshift measurements. And let us now denote redshift space measurements by the superscript $z$. The redshift space comoving position $\boldsymbol{x}^z$ of a particle located in $\boldsymbol{r}(\boldsymbol{q}) = a\boldsymbol{x}(\boldsymbol{q})$ is $\boldsymbol{x}^z = \dot{\boldsymbol{r}}/(aH)$ (with $H = \dot{a}/a$). The real space perturbative expansion (7) is then replaced by

$$\boldsymbol{x}^z = \boldsymbol{q} + [1 + f_1(t)]\, g_1(t) \tilde{\boldsymbol{\Psi}}^{(1)}(\boldsymbol{q})$$
$$+ [1 + f_2(t)]\, g_2(t) \tilde{\boldsymbol{\Psi}}^{(2)}(\boldsymbol{q}) + \mathcal{O}(\varepsilon^3), \tag{58}$$

where we have explicitely used the separability of $\boldsymbol{\Psi}^{(1)} = g_1(t)\tilde{\boldsymbol{\Psi}}^{(1)}(\boldsymbol{q})$ and $\boldsymbol{\Psi}^{(2)} = g_2(t)\tilde{\boldsymbol{\Psi}}^{(2)}(\boldsymbol{q})$ [eqs.(2.1) and (29)]. In the limit of an infinitely remote observer, say along the $r_3$-axis, the observed density constrast $\delta_z$ in comoving coordinates is simply $\delta^z(x_1, x_2, x_3^z)$, which amounts to approximate spherical coordinates by cartesian ones. All we have to do, then, is to replace everywhere in the calculation of $S_3$

$$\Psi_3^{(m)} = g_m \, \tilde{\Psi}_3^{(m)}$$

by

$$(1 + f_m) g_m \, \tilde{\Psi}_3^{(m)}$$

for $m = 1$ and $2$. We have,

$$\left\langle J^{(1)\,4} \right\rangle = [2 + (1 + f_1)^4]\, K + 3 \left\langle J^{(1)\,2} \right\rangle^2,$$

with

$$\left\langle J^{(1)\,2} \right\rangle = [2 + (1 + f_1)^2]\, \sigma^2.$$

This shows that in our so-called "infinite observer limit", we have $\left\langle \delta^2 \right\rangle^z = (1 + 2/3\, f_1 + 1/3\, f_1^2)\left\langle \delta^2 \right\rangle$ which slightly differs from Kaiser's calculation (who did a calculation in spherical coordinates instead of our rectangular ones). Indeed, we find that the redshift space variance is boosted by a factor $30/15$ for $\Omega = 1$, instead of his value of $28/15$.

The term $\langle J^{(1)\,2} J^{(2)} \rangle$ is now equal to

$$[2 + 4(1+f_1)^2 + (g_2/g_1^2)\,(2 + 4(1+f_1) + 2f_2 + f_1 f_2\, U)]\,\sigma^4$$

with

$$U\sigma^4 = 4\left\langle \Psi^{(1)}_{1,1}\Psi^{(1)}_{3,3}\Psi^{(2)}_{3,3}\right\rangle + (2+f_1)\left\langle \Psi^{(1)\,2}_{3,3}\Psi^{(2)}_{3,3}\right\rangle$$

containing all symmetry breaking terms. To evaluate $U$, we use discrete Fourier transforms, which we denote by hats, in a very large volume. It follows from (28) that

$$\hat{\Psi}^{(2)}_{3,3} = (g_2/g_1^2)\,(k_3^2/|\boldsymbol{k}|^2)\sum_{\boldsymbol{k}'}\sum_{i>j}$$

$$[k'_i(k_j - k'_j) - k'_j(k_i - k'_i)]\hat{\Psi}^{(1)}_i(\boldsymbol{k}')\hat{\Psi}^{(1)}_j(\boldsymbol{k} - \boldsymbol{k}')$$

(with $\Psi^{(2)}_{3,3}(\boldsymbol{x}) = \sum_{\boldsymbol{k}} \hat{\Psi}^{(2)}_{3,3}(\boldsymbol{k})\exp(i\boldsymbol{k}\cdot\boldsymbol{x})$). Since

$$\left\langle \hat{\Psi}^{(1)}_i(\boldsymbol{k})\hat{\Psi}^{(1)}_j(\boldsymbol{k}')\right\rangle = \delta_K(i-j)\delta(\boldsymbol{k} - \boldsymbol{k}')P(k),$$

i.e., is zero for $i \neq j$ ($\delta_K$ is the Kronecker symbol), or $\boldsymbol{k} \neq \boldsymbol{k}'$, it follows that $\left\langle \Psi^{(1)}_{i,i}\Psi^{(1)}_{3,3}\Psi^{(2)}_{3,3}\right\rangle$ is zero for $i = 3$ and is equal to

$$(g_2/g_1^2)\sum_{\boldsymbol{k},\boldsymbol{k}'} k_i^2 k'^2_3 (k_3 - k'_3)^2/(\boldsymbol{k} - \boldsymbol{k}')^2 P(k)P(k')$$

otherwise. Since $g_2 < 0$ (at late times when the growing mode is dominant), we have

$$\left\langle \Psi^{(1)}_{1,1}\Psi^{(1)}_{3,3}\Psi^{(2)}_{3,3}\right\rangle \leq 0;$$

similarly,

$$\left\langle \Psi^{(1)}_{1,1}\Psi^{(1)}_{3,3}\Psi^{(2)}_{2,2}\right\rangle \leq 0,$$

which implies

$$2\left\langle \Psi^{(1)}_{1,1}\Psi^{(1)}_{3,3}\Psi^{(2)}_{3,3}\right\rangle \geq \left\langle \Psi^{(1)}_{1,1}\Psi^{(1)}_{3,3}\nabla\boldsymbol{\Psi}^{(2)}\right\rangle = (g_2/g_1^2)\,\sigma^4.$$

As a result, $U$ is bounded, negative, and lies between 0 and $g_2/g_1^2$. It follows that

$$S_3^z = 6 - \frac{(2 + (1+f_1)^3)\,S}{\varepsilon\,(2 + (1+f_1)^2)^2\,\sigma^4}$$
$$- 2\,\frac{(2 + (1+f_1)^3)^2\,S^2}{(2 + (1+f_1)^2)^3\,\sigma^6}$$
$$+ 2\,\frac{(2 + (1+f_1)^4)\,K}{(2 + (1+f_1)^2)^2\,\sigma^4}$$
$$- 6\,\frac{1 + 2(1+f_1)^2 + (g_2/g_1^2)\,[3 + 2f_1 + f_2 + f_1 f_2 \mathcal{E}]}{[2 + (1+f_1)^2]^2} \quad (59)$$

with $1 \geq \mathcal{E} \geq 0$. Of course, we recover the real space result (57) if we set $f_1 = f_2 = 0$. On the other hand, if $\Omega = 1$, we have $f_1 = 1 = f_2/2$ (and $g_2/g_1^2 = -3/7$), which yields for Gausssian initial conditions $S_3 = (35 + \mathcal{E})/7$, while, for $\Omega = 0.1$, $S_3 \approx (34.5 + 0.4\mathcal{E})/7$ (to be compared with the value of $34/7$ in real space).

The formula (59) obtained above applies only in the limit of large volumes (like Kaiser's result), since we have taken the limit of an infinitely remote observer. A full calculation along the previous lines, but in spherical coordinates (and including the effect of smoothing) may be found in Hivon et al. (1994). In any case, equation (59) clearly shows that the ratio $S_3$ is, for gaussian initial conditions, nearly independent of the value of $\Omega$, nor is it affected by redshift space distortions. It is interesting to note, though, that in the non-gaussian case, the distortion might be rather large, for large enough S or K.

## 4. Comparison with other approximations for spherically symmetric perturbations

So far, we have mainly considered rigorous uses of the Lagrangian perturbative approach, for instance the derivation of a second-order quantity, the skewness of a PDF with the help of second order perturbation theory. Now, we examine to what extent the second-order theory brings improvement to Zeldovich approximation, when both are used as approximations to the real dynamics, i.e. outside of their rigorous validity range. In this section, we first explicitly derive Eulerian expressions from their Lagrangian counterparts (up to second order). Then we use the spherically symmetrical model to compare various perturbative approaches with the exact solution. In the next section, we shall compare to numerical simulations from Gaussian initial condition with a power-law power spectrum.

### 4.1. Lagrangian/Eulerian approach up to second order

Knowing the fastest growing solution up to second order in the Lagrangian formalism, we are now able to find its Eulerian counterpart. In this section, it is more convenient to slightly change the previous notations: we drop the $\varepsilon$ (or equivalently the displacement field is redefined to include the epsilons) and we use $g_{j,i} \equiv g_j(\tau = \tau_i) = g_j(t = t_i)$, $j = 1, 2$ (i.e. we do not impose $g_j(t = t_i) = 1$ anymore). The Eulerian initial conditions are generally specified by an initial density profile

$$\delta_i(\boldsymbol{x}) = \delta(\boldsymbol{x}, t_i) = \mathcal{O}(\varepsilon), \tag{60}$$

and an initial velocity profile. We assume, as before, that $\delta_i$ is dominated by the fastest growing mode. Let us also recall that, at lowest order in $\varepsilon$, the initial displacement is determined by

$$\nabla_q \cdot \boldsymbol{\Psi}(t_i, \boldsymbol{q}) \equiv \nabla_q \cdot \tilde{\boldsymbol{\Psi}}^{(1)}(\boldsymbol{q}) = -\delta_i(\mathbf{x}_i[\boldsymbol{q}]), \tag{61}$$

$$\nabla_q \times \tilde{\boldsymbol{\Psi}}^{(1)}(\boldsymbol{q}) = 0. \tag{62}$$

The mass conservation equation ($\delta = J^{-1} - 1$) yields

$$\delta(\boldsymbol{x}[\boldsymbol{q}]) = -\frac{g_1}{g_{1,i}}\nabla_q.\tilde{\boldsymbol{\Psi}}^{(1)} + \frac{g_1^2}{g_{1,i}^2}(\nabla_q.\tilde{\boldsymbol{\Psi}}^{(1)})^2$$
$$- \frac{g_1^2}{g_{1,i}^2}\left[1 + \frac{g_2}{g_1^2}\right]\nabla_q.\tilde{\boldsymbol{\Psi}}^{(2)} + \mathcal{O}(\varepsilon^3), \quad (63)$$

where $\tilde{\boldsymbol{\Psi}}^{(2)}(\boldsymbol{q})$ is the curl-free quantity verifying

$$\nabla_q.\tilde{\boldsymbol{\Psi}}^{(2)} = \frac{1}{2}\sum_{l\neq m}\left(\tilde{\boldsymbol{\Psi}}^{(1)}_{l,l}\tilde{\boldsymbol{\Psi}}^{(1)}_{m,m} - \tilde{\boldsymbol{\Psi}}^{(1)}_{l,m}\tilde{\boldsymbol{\Psi}}^{(1)}_{m,l}\right). \quad (64)$$

It is then useful to introduce the Eulerian displacement $\tilde{\boldsymbol{\Psi}}^{(1e)}(\boldsymbol{x})$, defined by

$$\tilde{\boldsymbol{\Psi}}^{(1e)}(\boldsymbol{x}) \equiv \tilde{\boldsymbol{\Psi}}^{(1)}(\boldsymbol{q} = \boldsymbol{x}). \quad (65)$$

By construction, this displacement is curl-free in Eulerian space, i.e. $\nabla_x \times \tilde{\boldsymbol{\Psi}}^{(1e)} = 0$. It is simply related to the initial density contrast by $\nabla_x.\tilde{\boldsymbol{\Psi}}^{(1e)} = -\delta_i(\boldsymbol{x})$. We can write

$$\tilde{\boldsymbol{\Psi}}^{(1)}(\boldsymbol{q}) = \tilde{\boldsymbol{\Psi}}^{(1e)}(\boldsymbol{x}) - \frac{g_1}{g_{1,i}}\overline{\overline{\nabla}}_x\tilde{\boldsymbol{\Psi}}^{(1e)}.\tilde{\boldsymbol{\Psi}}^{(1e)} + \mathcal{O}(\varepsilon^2), \quad (66)$$

$$\nabla_q.\tilde{\boldsymbol{\Psi}}^{(1)} = \nabla_x.\tilde{\boldsymbol{\Psi}}^{(1e)} - \frac{g_1}{g_{1,i}}\left\{\nabla_x(\nabla_x.\tilde{\boldsymbol{\Psi}}^{(1e)})\right\}\tilde{\boldsymbol{\Psi}}^{(1e)} + \mathcal{O}(\varepsilon^2). \quad (67)$$

The Eulerian density contrast $\delta(\boldsymbol{x})$ then is:

$$\delta(\mathbf{x}) = -\frac{g_1}{g_{1,i}}\nabla_x.\tilde{\boldsymbol{\Psi}}^{(1e)}$$
$$+ \frac{g_1^2}{g_{1,i}^2}[(\nabla_x.\tilde{\boldsymbol{\Psi}}^{(1e)})^2 + \{\nabla_x(\nabla_x.\tilde{\boldsymbol{\Psi}}^{(1e)})\}.\tilde{\boldsymbol{\Psi}}^{(1e)}]$$
$$- \frac{g_1^2}{g_{1,i}^2}\left[1 + \frac{g_2}{g_1^2}\right]\nabla_x.\tilde{\boldsymbol{\Psi}}^{(2e)} + \mathcal{O}(\varepsilon^3). \quad (68)$$

The quantity $\tilde{\boldsymbol{\Psi}}^{(2e)}(\boldsymbol{x})$ is curl-free and verifies

$$\nabla_x.\tilde{\boldsymbol{\Psi}}^{(2e)} = \frac{1}{2}\sum_{l\neq m}\left(\tilde{\boldsymbol{\Psi}}^{(1e)}_{l,l}\tilde{\boldsymbol{\Psi}}^{(1e)}_{m,m} - \tilde{\boldsymbol{\Psi}}^{(1e)}_{l,m}\tilde{\boldsymbol{\Psi}}^{(1e)}_{m,l}\right). \quad (69)$$

A similar transformation can be made for the velocity field. In the Lagrangian approach, the peculiar velocity of an element of matter is written as

$$\frac{\mathbf{v}(\boldsymbol{q},t)}{a} = \frac{\dot{g_1}}{g_{1,i}}\tilde{\boldsymbol{\Psi}}^{(1)}(\boldsymbol{q}) + \frac{\dot{g_2}}{g_{1,i}^2}\tilde{\boldsymbol{\Psi}}^{(2)}(\boldsymbol{q}) + \mathcal{O}(\varepsilon^3). \quad (70)$$

This leads in Eulerian coordinates to

$$\frac{\mathbf{v}(\boldsymbol{x},t)}{a} = \frac{\dot{g_1}}{g_{1,i}}\tilde{\boldsymbol{\Psi}}^{(1e)}(\boldsymbol{x})$$
$$+ \frac{\dot{g_2}}{g_{1,i}^2}\tilde{\boldsymbol{\Psi}}^{(2e)}(\boldsymbol{x}) - \frac{\dot{g_1}g_1}{g_{1,i}^2}\overline{\overline{\nabla}}_x\tilde{\boldsymbol{\Psi}}^{(1e)}.\tilde{\boldsymbol{\Psi}}^{(1e)} + \mathcal{O}(\varepsilon^3). (71)$$

For further reference, let us recall that Zeldovich approximation is nothing else than the fastest linear growing mode of the Lagrangian perturbative expansion (7). What we will call Eulerian linear theory corresponds to the first term in right hand of Eqs. (68), (71).

Strictly speaking, the second order expressions given above are valid only if

$$g_1/g_{1,i} \gg 1, \quad (72)$$

$$\varepsilon g_1/g_{1,i} \ll 1. \quad (73)$$

The first condition insures that transient modes and subdominant modes are negligible. The second condition is necessary for the perturbative solution to be valid. However, as we will see in the next paragraph when we consider the spherically symmetrical model, the condition (73) may be relaxed in the Lagrangian case, as far as the density contrast is concerned. Indeed, mass conservation is by construction verified in the Lagrangian approach. It explains why Zeldovich approximation, although a first order (linear) solution, provides a very good qualitative description of the density distribution in pancake models. In the Eulerian case, though, the condition (73) has to be obeyed, otherwise the approximation (68) becomes completely wrong. With regard to velocity fields, one can expect the Eulerian perturbative approach to be almost as accurate as the Lagrangian one at the same order, since mass conservation is less important. For example, roughly speaking, expressions (70) and (71) are equivalent when taken at first order, if (70) is evaluated at $\boldsymbol{x}' = \boldsymbol{q}$.

### 4.2. Example: the spherical model

The spherical model can be exactly solved analytically before shell-crossing. It thus provides a good test for various perturbative models. The aim of this paragraph is to compare different approximations with the exact solution. Firstly, we consider the case of the top hat model, which is equivalent to look at the center of the perturbation. The density contrast will be studied, as well as the divergence of the velocity field. Then, the density contrast and the velocity field of a given profile will be analyzed as functions of the radial coordinate $r$. The perturbative solutions studied here are the standard Eulerian linear approximation, the Zeldovich approximation, the Eulerian Second order approximation and the Lagrangian second order approximation. Details of the solutions for the evolution in the various approximations are given in appendix A. In Section 4.2.1, we also consider Bernardeau's model (1992). This model provides an exact relationship between the density contrast and the divergence of the velocity field in the weakly nonlinear limit (when the variance $\langle\delta\rangle^2$ of the distribution is infinitely small). It is based on a statistical approach, so it may be somewhat meaningless to use it here, since we consider a particular symmetry. However, it will be interesting to notice that the calculation of Bernardeau provides a very good fit of the relation between $\nabla.\mathbf{v}$ and $\delta$ for the spherical top hat model, which is rather useful when one wants to use it as a toy model, since exact solutions have rather complicated expressions.

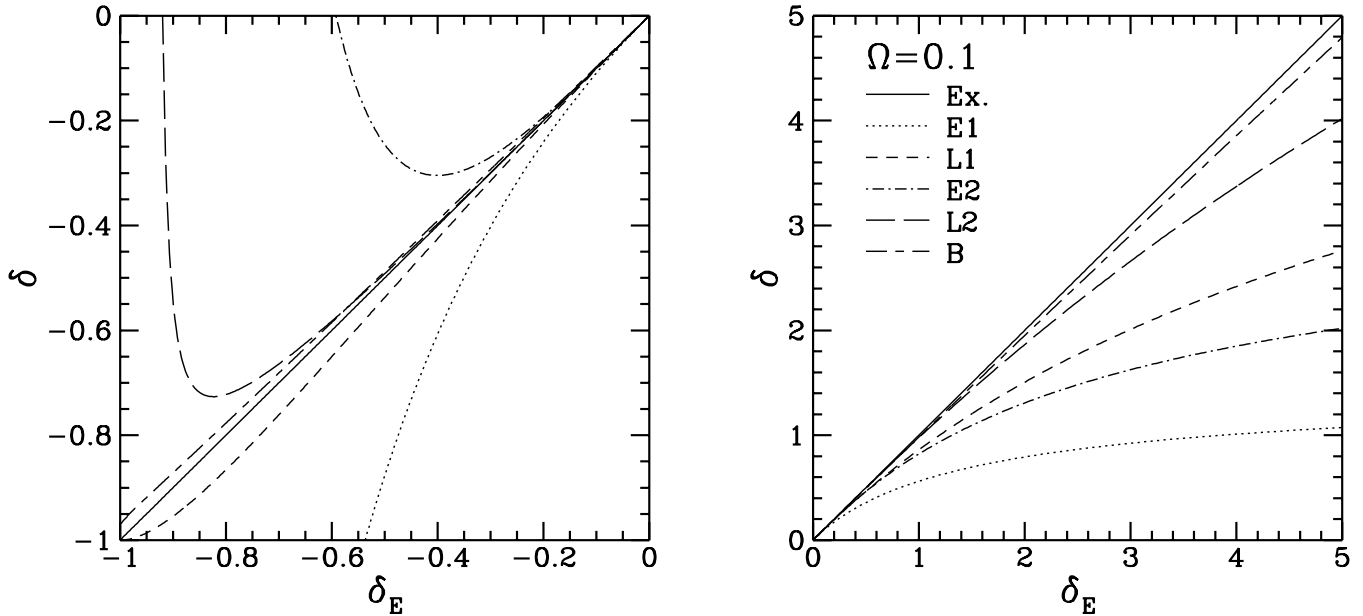

**Fig. 6.** Computed density contrast $\delta$ as a function of the exact solution $\delta_E$ for various approximations for a spherical top hat perturbation ($\Omega = 0.1$). The left panel and the right panel respectively correspond to $\delta_E \leq 0$ and $\delta_E \geq 0$. The abbreviated labels in the right panel correspond to the following models: "Ex." stands for the exact solution, "E1" for Eulerian linear theory, "E2" for Eulerian second order perturbation theory, "L1" for Zeldovich approximation, "L2" for Lagrangian second order perturbation theory, and "B" for Bernardeau's model.

In the following, we assume that shell-crossing has not taken place. Results are given for $\Omega = 0.1$ (and $\Lambda = 0$), but the qualitative behavior of the solutions does not change significantly with $\Omega$ and the conclusions given here can be generalized for any reasonable value of $\Omega$ ($0.05 \lesssim \Omega \lesssim 3$).

### 4.2.1. The top hat model

Let us consider an initial density profile given by

$$\begin{cases} \delta(x_i, t_i) = \delta_i, & x_i \leq x_{1,i}, \\ \delta(x_i, t_i) = 0, & x_i > x_{1,i}. \end{cases} \quad (74)$$

According to Birkoff's theorem, the evolution of a Lagrangian sphere of initial radius $x_i < x_{1,i}$ depends only on its contents. Then the matter distribution inside the sphere of radius $x_1(t)$, with $x_1(t_i) \equiv x_{1,i}$ remains homogeneous while the system is evolving. For $x \leq x_1(t)$, the density contrast $\delta(x,t)$ is thus a function of time and $\delta_i$. This reasoning can be generalized to any initial density profile $\delta(x_i, t_i)$, when one considers the center of the perturbation: $\delta(0,t)$ depends only on $\delta(0,t_i)$ and $t$.

Figure 6 gives the density contrast $\delta$ as a function of the exact solution $\delta_E$ for various perturbative models and for Bernardeau's model. As expected, the Lagrangian approach seems to be much more efficient than the Eulerian one. Even Zeldovich approximation, which is only valid at first order, is better than the Eulerian second order approximation (hereafter L2). Note that at second order, arbitrarily underdense regions cannot be obtained. In the Eulerian case, the largest underdensity that can be reached is

$$\delta_{\min,\text{E2}} \simeq -\frac{3}{28(14 + 3\Omega^{-2/63})} \simeq -0.3. \qquad (75)$$

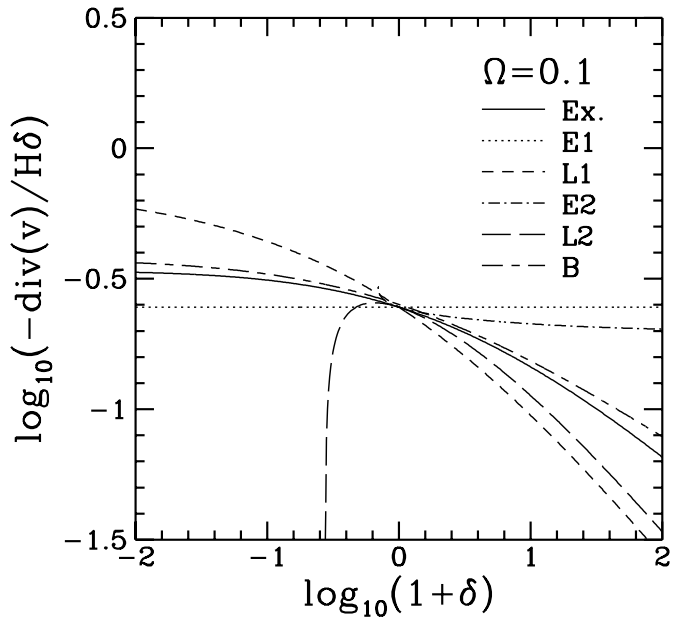

**Fig. 7.** The quantity $-\nabla_x \cdot \mathbf{v}/(aH\delta)$ as a function of $\delta$, in logarithmic coordinates, for various approximations in the spherical top hat model ($\Omega = 0.1$). Notations are the same as those used in Fig. 6. The Lagrangian second order approach (L2) and the Eulerian second order approach (E2) cannot produce arbitrarily small density [see Fig. 6, Eqs. (75), (76)], which explains the discontinued line for E2 and the vertical asymptote for L2.

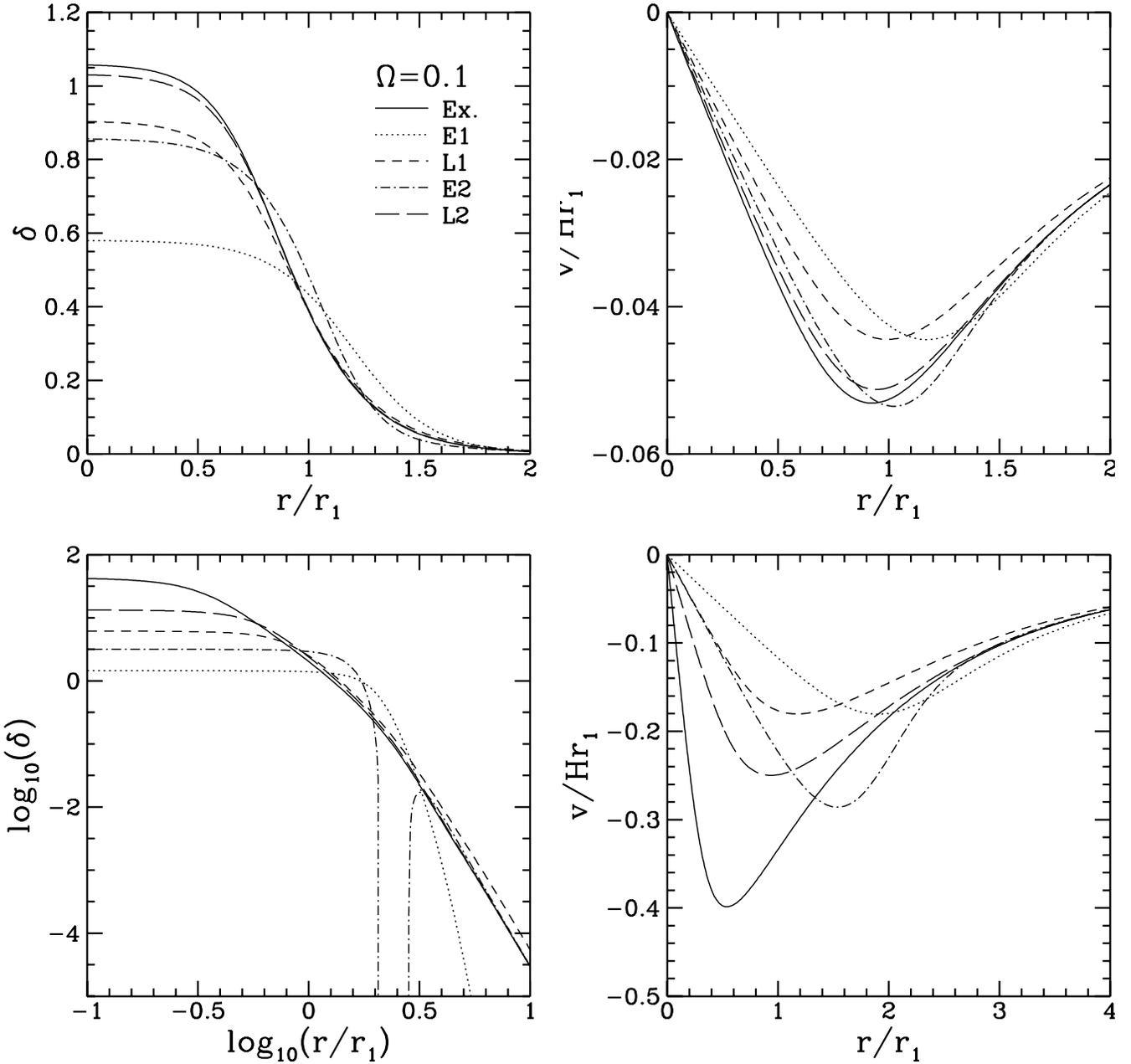

**Fig. 8.** The density contrast (left panels) and the velocity field (right panels) versus the distance to the center of the halo for two initial values of $\overline{\delta}$ (see text). The top and bottom panels respectively correspond to $\overline{\delta} = 0.002$ and $\overline{\delta} = 0.005$. For a better legibility, the bottom left graph is in logarithmic coordinates.

For Lagrangian second order perturbation theory, we have instead

$$\delta_{\min,\text{L2}} \simeq \left(1 + \frac{7}{12}\Omega^{2/63}\right)^{-3} - 1 \simeq -0.75, \qquad (76)$$

which is of course much better. In the regime $-0.8 \lesssim \delta \lesssim 3$, the accuracy of the Lagrangian second order approximation is better than 7%. Although it considers a statistical average and not a single object, Bernardeau's model is quite good.

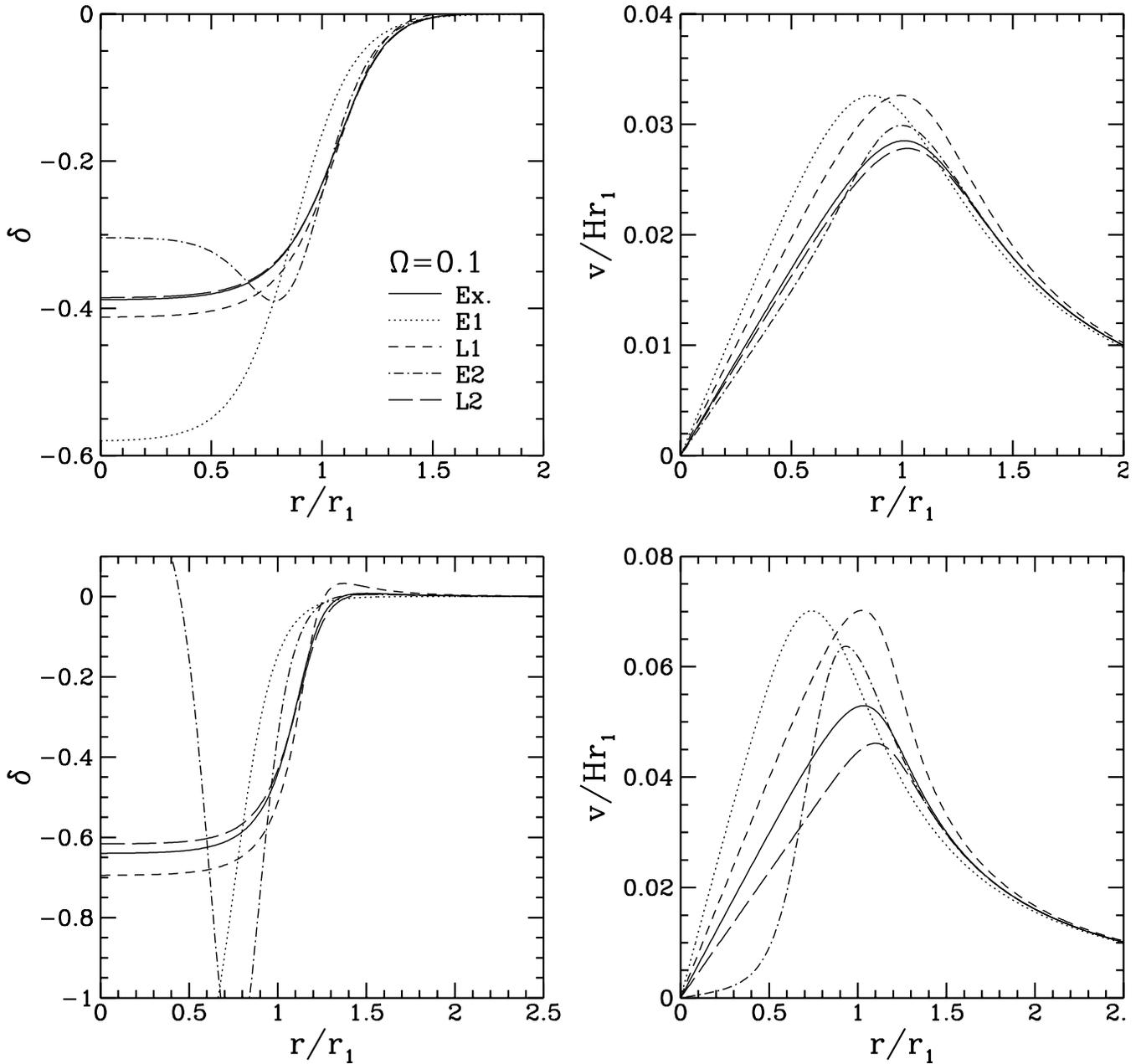

**Fig. 9.** Same as in Fig. 8, but for a void. Even for a moderate final density contrast ($\delta \sim -0.4$), the Eulerian second order approach gives bad results.

Figure 7 is similar to Fig. 6, but it gives the quantity $\nabla_{\mathbf{x}}.\mathbf{v}/(aH\delta)$ as a function of $\delta$. As in Fig. 6, the model of Bernardeau gives a very good approximation of the exact solution. One can see that a second order solution is needed to get the appropriate slope of $\nabla_x.\mathbf{v}/(aH\delta)$ in the vicinity of $\delta = 0$, as expected. Indeed, this slope corresponds to the first correction to the horizontal dotted line, which refers to the Eulerian linear theory. In this case, Zel-dovich approximation is not really better than the Eulerian linear theory. This is not surprising, as already argued in Sect. 4.1. Similar conclusions can be reached when one considers second order approximations. The discontinued line in the Eulerian second order case and the vertical asymptote in the Lagrangian second order case both reflect the existence of a maximum underdensity that can be reached in these approximations.

### 4.2.2. Study of a given profile

The spherical top hat model is quite extreme. On can indeed expect that perturbative models fail in approaching the exact solution at the center of the fluctuation. As we shall see through an example, for a given initial profile $\delta_i(r_i)$, the Lagrangian second order approximation is able to describe density contrasts as large as ten, if one considers regions far enough from the center of the fluctuation.

Following Martel & Freudling (1991, hereafter MF), we take as initial profile

$$\delta_i(r_i) = \frac{\overline{\delta}}{2} \left[ 1 - \text{th} \left( \frac{r_i - r_{1,i}}{\xi r_{1,i}} \right) \right]. \qquad (77)$$

The typical radius of the initial perturbation is thus given by $r_{1,i}$, while $\xi$ parameterizes the width of the transition between the regime $\delta_i(r_i) \simeq \overline{\delta}$ and the regime $\delta_i(r_i) \simeq 0$. We choose, as MF,

$$r_{1,i} = 1, \quad \xi = 0.3, \quad a/a_i = 1500. \qquad (78)$$

Figures 8 and 9 give for a spherical halo and a void, the density contrast and the velocity field as functions of $r/r_1(t)$. The function $r_1(t)$ is the Lagrangian position of points initially at $r_{1,i}$, or in other words, the typical radius of the fluctuation at $t$. The peculiar velocity $v$ is given in units of the Hubble flow for $r = r_1(t)$. We take

$$|\overline{\delta}_1| = 0.002, \quad |\overline{\delta}_2| = 0.005, \qquad (79)$$

to test the various approximations: $|\overline{\delta}_1|$ is such that the final value of $\delta$ is moderate, about unity for a halo and about $-0.4$ for a void; $|\overline{\delta}_2|$ leads to large density contrasts, about 40 for a halo and about $-0.6$ for a void. The calculations are assuming $\Omega_0 = 0.1$, but the results are quite similar for different $\Omega_0$, although one has to decrease the value of $\overline{\delta}$ when $\Omega_0$ increases, in order to get the same final $\delta$. The values of $\Omega_0$ and $a/a_i$ choosen here give $g_1/g_{1,i} \simeq 290 \gg 1$, so the validity condition (72) is verified. But condition (73) is not verified, since we get $\varepsilon g_1/g_{1,i} \simeq 0.58$ for $|\overline{\delta}| = |\overline{\delta}_1|$ and $\varepsilon g_1/g_{1,i} \simeq 1.45$ for $|\overline{\delta}| = |\overline{\delta}_2|$. This is outside the appropriate regime for Eulerian second order perturbation theory at least for $|\overline{\delta}| = |\overline{\delta}_2|$.

Figs. 8 and 9 confirm the results of the previous paragraph, in terms of relative accuracy of the various approximations for *moderate* final density contrasts:

**density contrast** :
Lagrangian second order > Zeldovich $\gtrsim$      (80)
Eulerian second order     > Eulerian linear theory,

**Velocity field** :
Lagrangian second order $\sim$ Eulerian second order > (81)
Zeldovich                    $\sim$ Eulerian linear theory.

For *large* final density contrasts, the Eulerian approach becomes particularly inefficient, while the advantage of the Lagrangian approach is much less pronounced for the velocity field.

Overall, the second order Lagrangian approach gives, for moderate final $\delta$, an excellent approximation of the density contrast and the velocity field. It appears to correctly reproduce density contrasts as large as ten. Thus these comparisons suggest that by releasing condition (73), i.e. by extrapolating the perturbative Lagrangian solutions and using them as approximations to the real dynamics, one gets a good idea of the behavior of the system in the non-linear regime.

## 5. Comparisons with a Numerical simulation with Gaussian initial perturbations with a power spectrum $\propto k^{-2}$

In order to see whether the conclusions reached in the spherical case hold in a more realistic setup, we now study the evolution of three-dimensional, Gaussian, initial perturbations with a power spectrum $\propto k^{-2}$. A similar study (with similar conclusions), but for truncated power spectra may be found in Melott et al. (1994). We compare the evolution of $64^3$ particles in a $128^3$ PM numerical simulation with the evolution computed by using the first (Zeldovich, or L1) and second order (L2) Lagrangian approximations. Periodic boundary conditions are assumed. All Eulerian quantities such as the density and the velocity fields are computed on a $64^3$ grid.

### 5.1. A Slice through the data

The figure 10 shows the particles positions in the same thin slice of the box, $L_{box}/64$ thick, when the simulation (left column), the L1 (Zeldovich, middle column), and L2 (right column) approximations are used to compute the evolution of the same initial conditions (hereafter IC). The rows from top to bottom show the positions after an expansion by a factor $a = 4, 8,$ and 16, respectively.

Of course, the approximations fail to describe the inner structure of regions where shell-crossing has occurred, and it is interesting to rather compare the resulting smoothed density fields. The first three rows of figure 11 display the density contours corresponding to the slices of figure 10. As is already well-known, Zeldovich approximation, despite its simplicity, does very well at describing the general large-scale texture of the density field. We also note that at this purely visual level, L2 does not appear to bring very marked improvements over L1.

In both approximations, the trajectories are entirely determined by the initial local velocity (L1) and also the acceleration (L2) field. Thus, contrary to the real case, if small scale variations are initially present, they will get amplified and transferred to much larger scales (even more so for L2, see figure 1), since the recall force of a forming non-linear clump is absent. Kofman et al. (1992) first proposed to truncate the initial power spectrum of its small scale power to improve the results of Zeldovich approximation. And indeed Coles et al. (1993) and Melott et al. (1993) showed that this smoothing of the initial conditions, at a scale close to the scale of non-linearity when

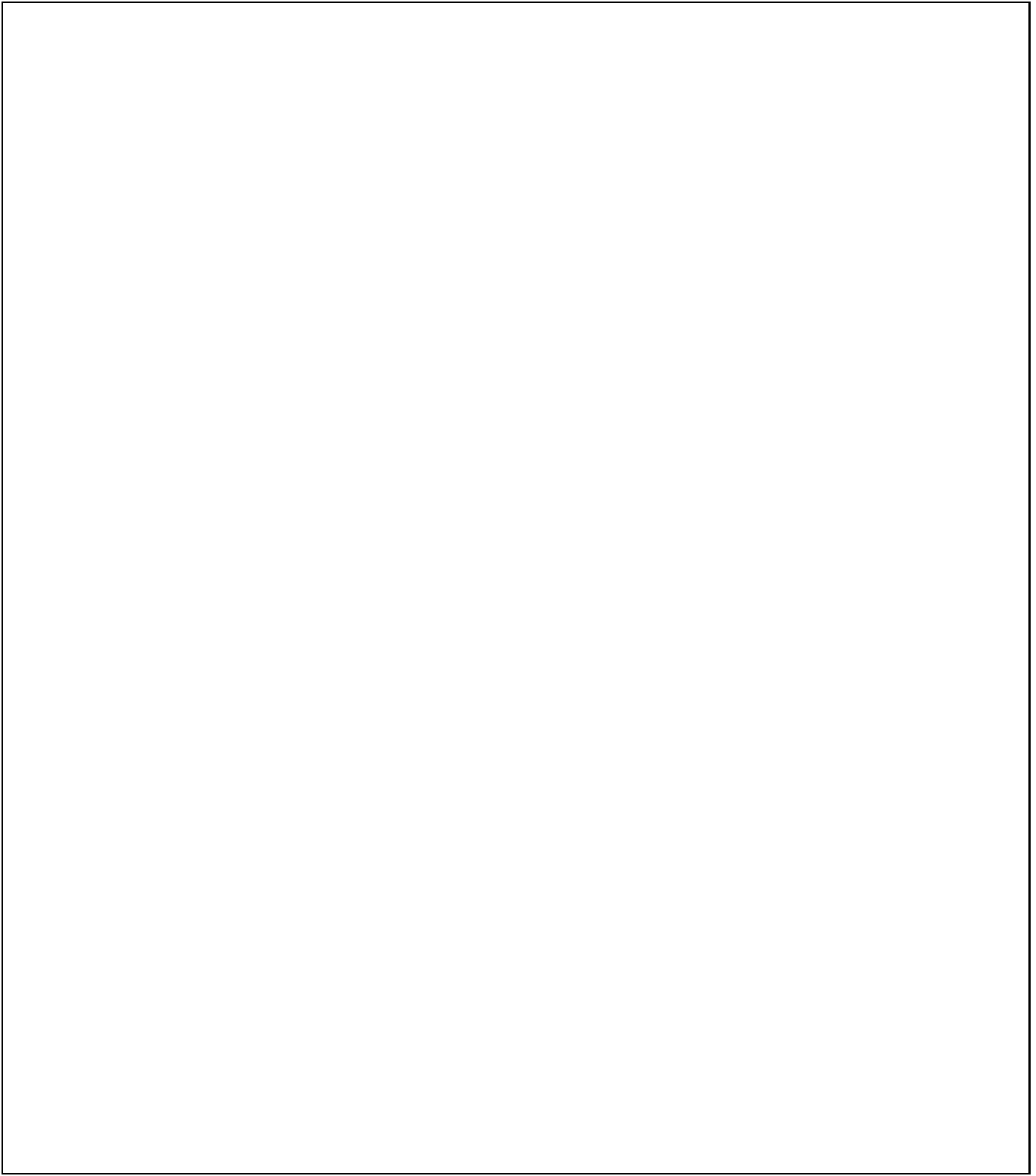

**Fig. 10.** Slices $L_{box}/50$ thick of a model evolved from $n = -2$ gaussian initial conditions with respectively a PM numerical simulation (left column), Zeldovich approximation (middle column) and the Lagrangian second order approximation (right column). The rows from top to bottom correspond to $a = 4$, 8, and 16.

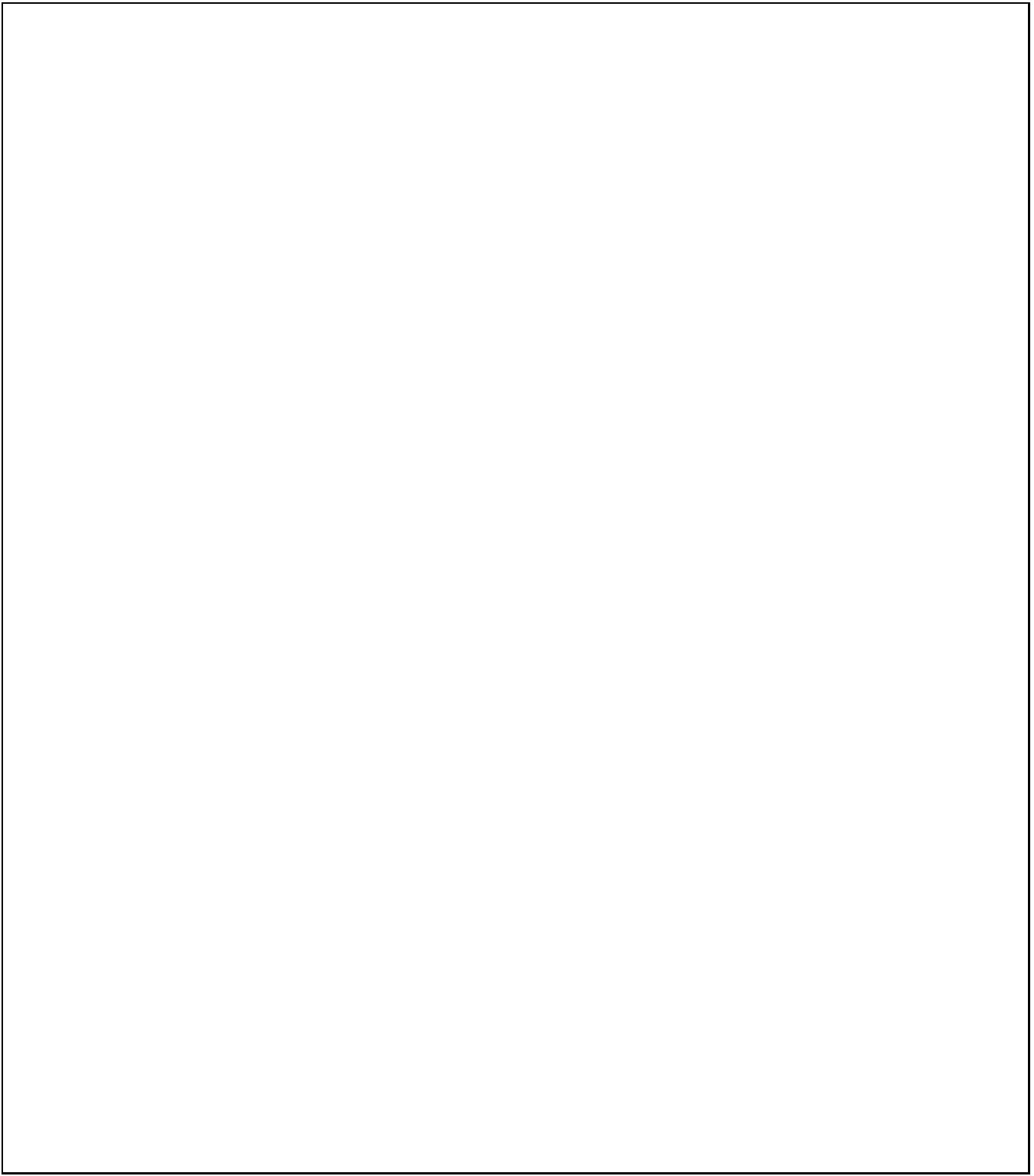

**Fig. 11.** The tree top rows show the density contours for the same slices than in the previous figure (i.e. PM, L1, and L2 are from left to right, and $a =$ 4, 8, and 16, from top to the third row). The left figure of the bottom row show the computed evolution of the variance (in cubic cells of size $1/64$), when the initial conditions are smoothed with Gaussian filters of various scale $\ell_s$. The adjacent figures show the corresponding $a = 16$ density contours for the L1 and L2 case, when an initial smoothing has applied at a scale $\ell_s = 1/64$, to be compared with the unsmoothed case directly above.

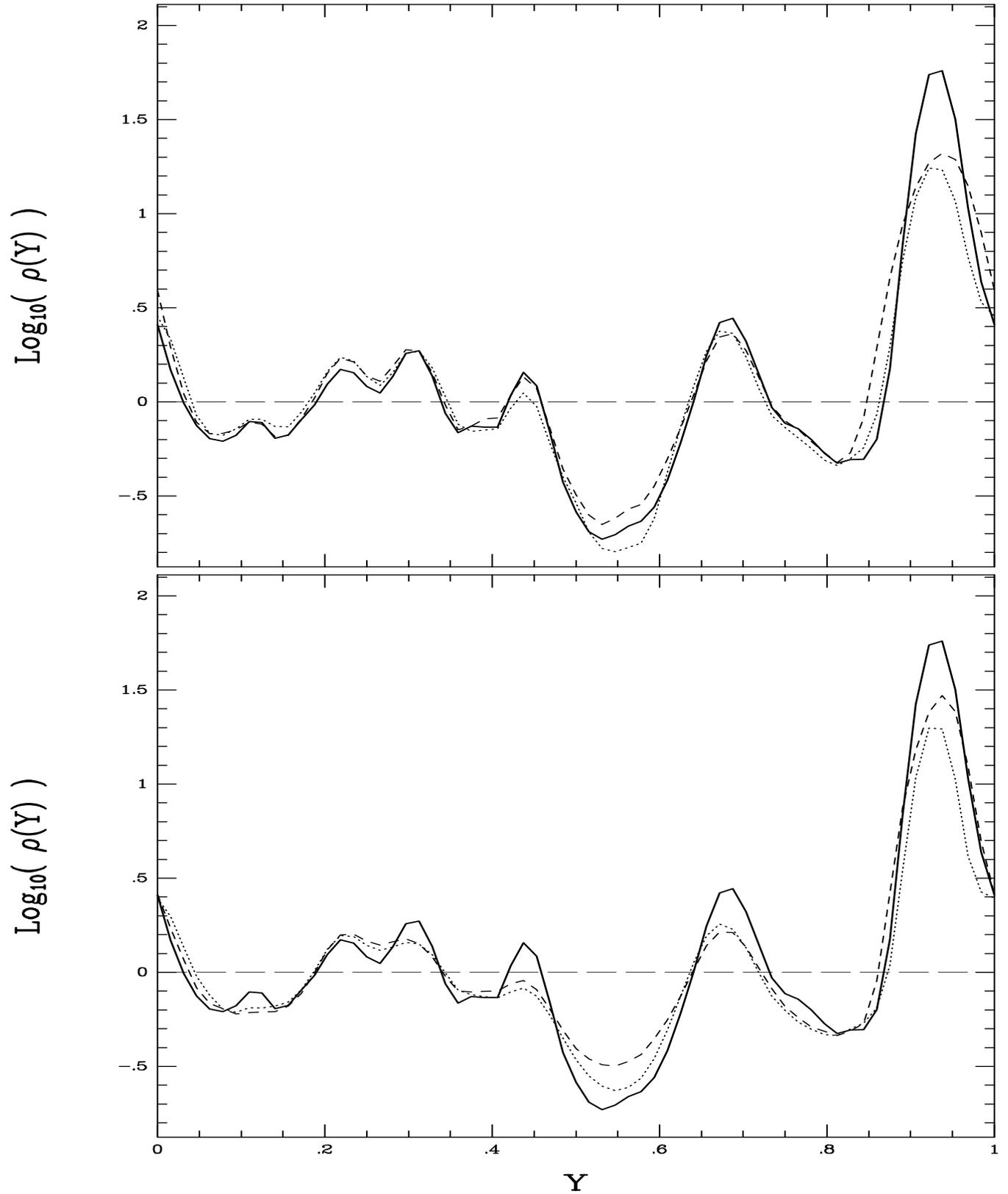

**Fig. 12.** Density measured along a one-dimensional scan in a box evolved with a PM simulation (solid), and by using the L1 and L2 approximations (dots and dashes respectively) at $a=16$. In both cases, the density has been obtained by smoothing the data with a gaussian filter with $\ell_s = L_{box}/64$ (see text). The top panel compares the results for identical initial conditions, while on the bottom the initial conditions for the L1 and L2 approximations have been smoothed with the same $\ell_s = L_{box}/64$ gaussian filter used at $a=16$.

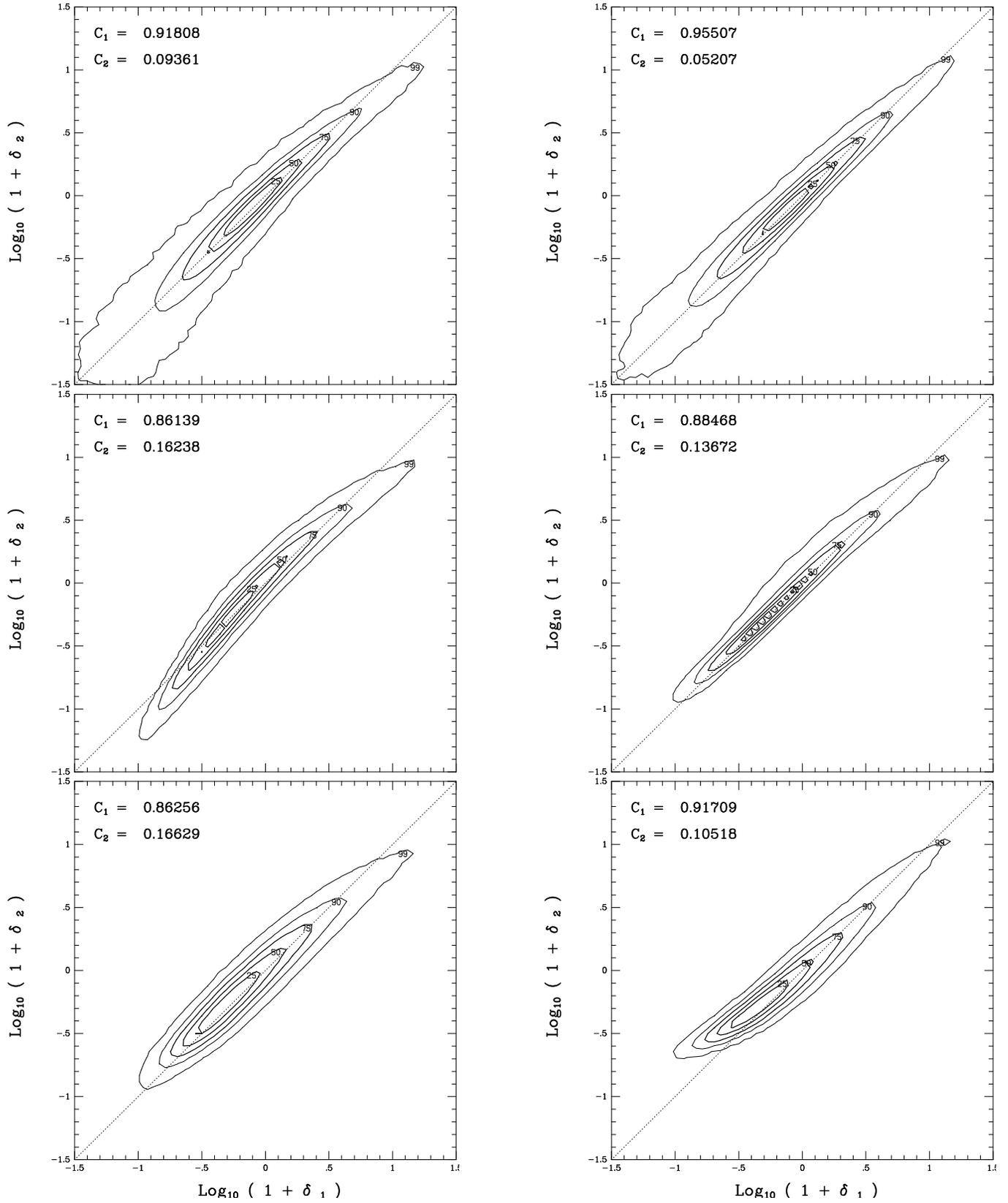

**Fig. 13.** Approximate density contrasts, $\delta_2$, obtained by L1 (left) and L2 (right) versus the PM one, $\delta_1$. The top panel is for $a = 8$. . The middle and bottom panels are at $a = 16$ and correspond respectively to unsmoothed and smoothed IC. See text for details.

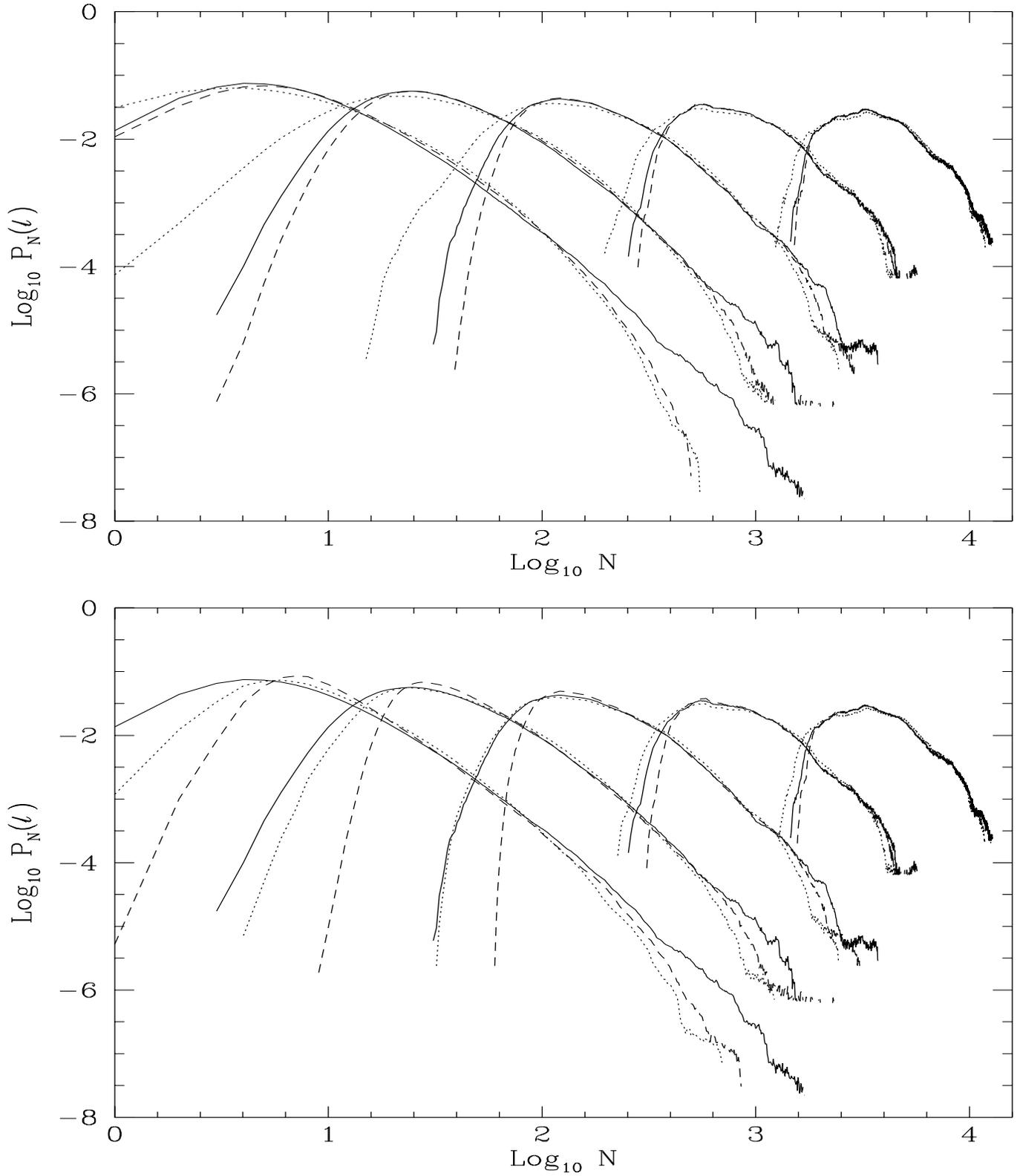

**Fig. 14.** The CPDF $P_N(\ell)$ at $a = 16$, for a PM, L1, and L2 "mover" (solid, dots, and dashes respectively). From left to right, the curves are for the scale $\log_{10} \ell$=-1.6, -1.4, -1.2, -1.0, -0.8. The corresponding variances (see eq.(85)) measured in the PM simulation are $\langle \delta^2 \rangle \simeq 3.3$, 1.6, 1/1.26, 1/2.5, 1/5. Note that for a greater legibility, each curve has been offset as compared to its left neighbor by +0.5 on the y-axis.

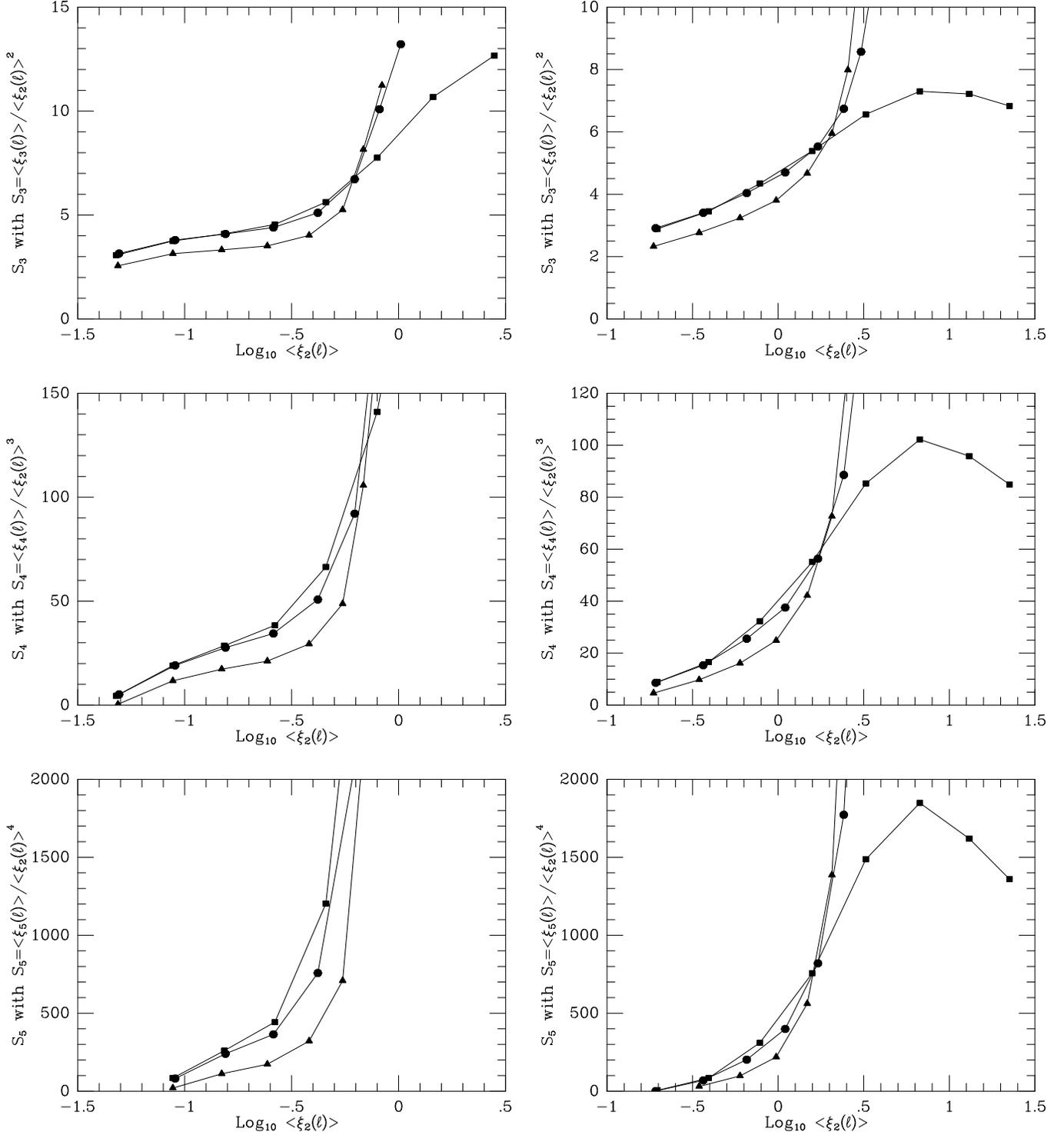

**Fig. 15.** The third, forth and fifth normalized moments of the density distribution for $a = 8$ (left) and $a = 16$ with smoothed initial conditions for approximations (right). The squares correspond to the simulation results, while triangles and circles denote the results obtained using respectively Zeldovich approximation and L2.

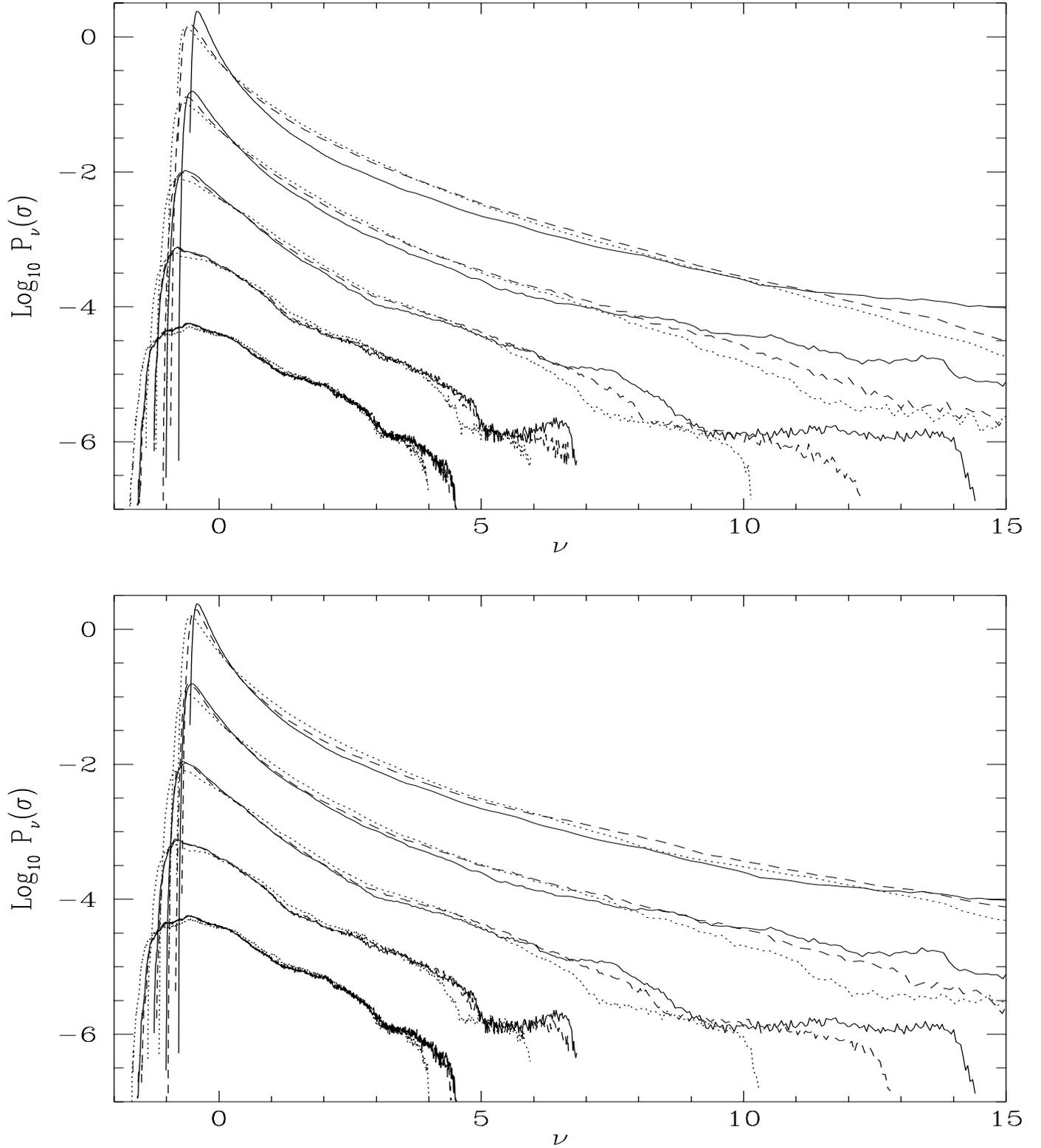

**Fig. 16.** Distribution function $P_\nu(\sigma)$ at $a = 16$, for a PM, L1, and L2 "mover" (solid, dots, and dashes respectively). The top pannel shows a comparison for unsmoothed intial conditions, and the bottom one for smoothed ones (with a gaussian filter of size $\ell_s = L_{box}/64$). In each panel, the curves from top to bottom, are for the following $\sigma^2_{PM}$ of the PM simulation : 3.3, 1.6, 1/1.26, 1/2.5, 1/5. For a greater legibility, each curve has been offset as compared to its neighbor above by -1 on the y-axis.

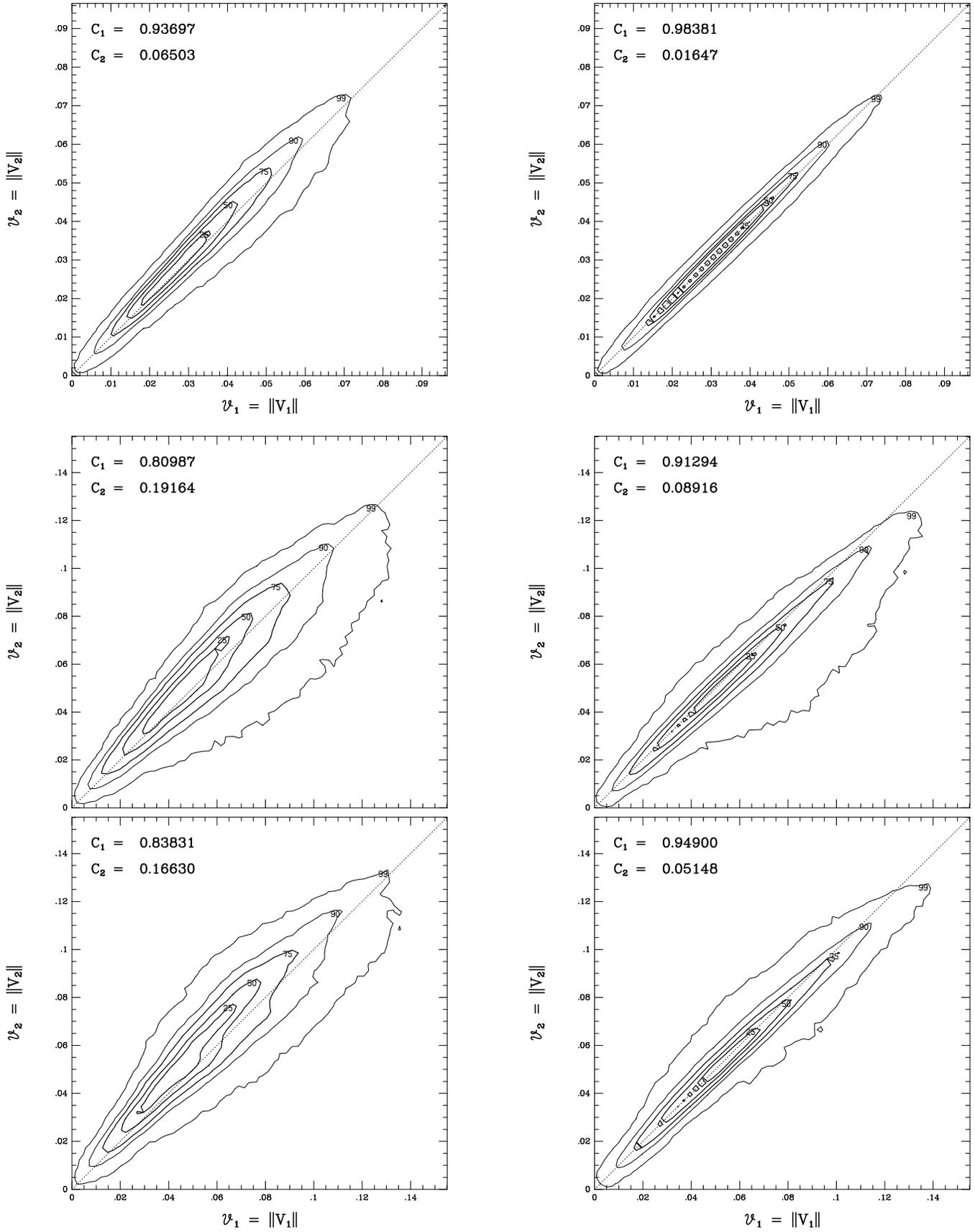

**Fig. 17.** Same as figure 13, but for the modulus of the velocity. L1 is on the left, L2 on the right. The top row is at $a=8$ and is not smoothed. The bottom rows corresponds to a latter time, $a=16$, when the measurements are performed after smoothing of the final velocity field, in the case of unsmoothed (middle) and smoothed (bottom) initial conditions.

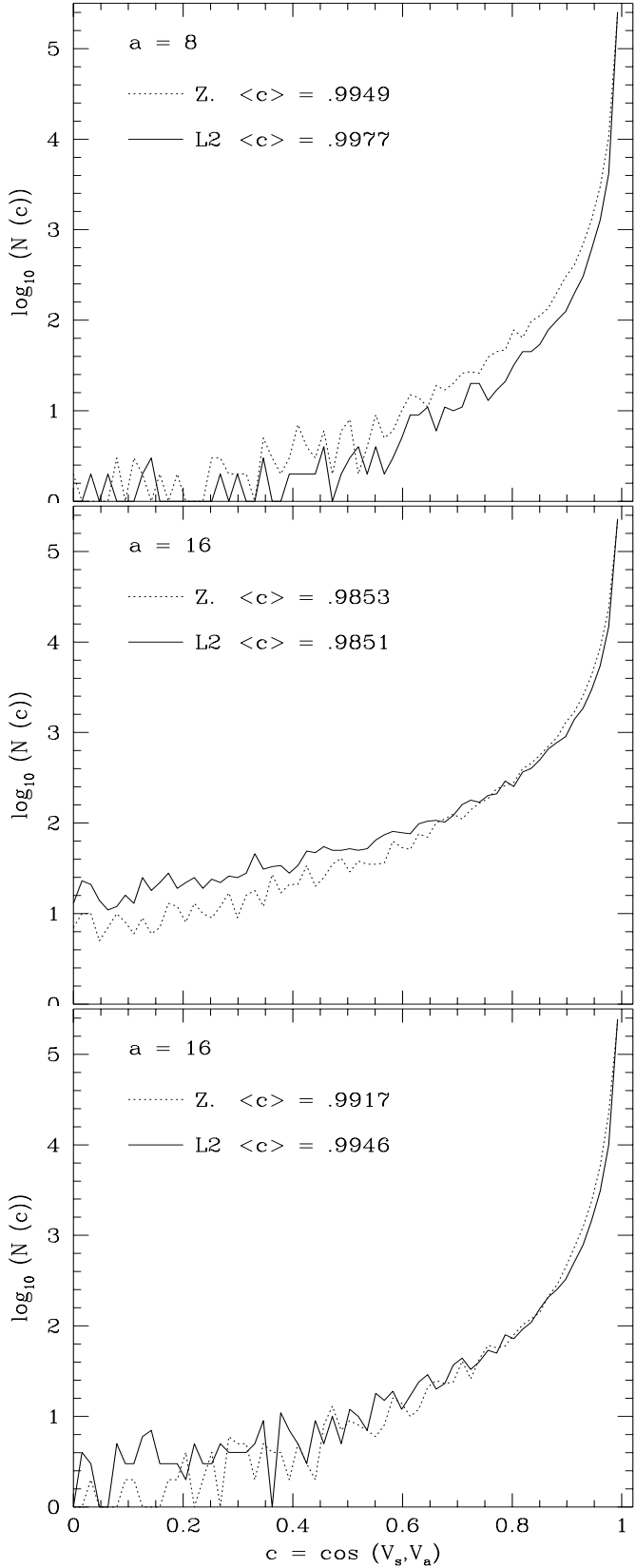

**Fig. 18.** Distribution of angles between the true velocity, and the one computed using L1 (dots), or L2 (solid). The top plot shows the result (for unsmoothed IC) at $a = 8$. The other plots show the result after smoothing of the final velocity field, at $a = 16$, for unsmoothed (middle) and smoothed IC (bottom).

the comparison is made, improves dramatically the predictions of Zeldovich approximation.

One may think of several ways of choosing the exact value of the scale at which the initial data is filtered. One would like to choose the smallest scale that still prevents, at the time the comparison is made, the unphysical "blow-up" of small scale structures. One way to do this is to choose that scale such the variance of the smoothed initial field is maximal at that time. The left panel of the bottom row of figure 11 shows the time evolution of the variance for various initial gaussian smoothing scales. It suggests that at $a = 8$, no smoothing should be necessary. And indeed we found no improvement in all the tests we considered when we used an initial smoothing with a filter of size $\ell_s = L_{box}/64$ (i.e. the filter $W(x)$ is $\propto \exp(-x^2/2\ell_s)$). But the same figure suggests that this smoothing scale should be appropriate at $a = 16$. The adjacent middle and right column figures of the bottom row display for reference the density contours at $a = 16$ for the L1 and L2 case, when this initial smoothing has been applied before computing the particles trajectories. In the following, we shall refer to these cases as L1S and L2S.

This visual comparison suggests that indeed an initial smoothing leads to a better, more contrasted, description of structures at late stages. We now proceed to a finer analysis of the computed evolution.

## 5.2. Density scans along a line

The figure 12 shows the density along a line, for the different approximations at $a = 16$. This line was chosen to go through the densest structure present in the simulation. The density was obtained by first affecting the particles to a $64^3$ grid using a cloud-in-cell interpolation, and then convolving the resulting field with a gaussian filter of $\ell_s = 1/64$. The top panel compares the results for identical initial conditions, evolved with the PM (Solid), L1 (dots) and L2 (dashes) movers. The bottom panels compares the L1S and L2S density scans when the initial conditions for the L1 and L2 approximations have been smoothed with the same $\ell_s = L_{box}/64$ gaussian filter used to obtain the final field, at $a=16$.

For relatively weak density contrasts, $3 \gtrsim \rho \gtrsim 1/3$, L2 is marginally better than L1, and both are much better than L1S and L2S. In a strongly underdense regions (at $Y \simeq .55$), L1 overestimates the density contrast, while L2 underestimates it. For smooth initial conditions, both L1S and L2S underestimate the real density contrast, but of course L1S is much better than L2S, since L1 was initially overestimating it. In a strongly clustered region (at $Y \simeq .95$), both L1S and L2S do much better than L1 and L2 (i.e. both the peak width and height are closer to the PM values), and L2S clearly surpasses L1S in terms of peak width and height.

These scans therefore suggest that smoothing is only appropriate for improving results in very dense regions, for which L2 appears superior to L1, the reverse being true in the underdense ones. These findings thus tend to confirm the results obtained in the spherical case (see Eq. (76)).

### 5.3. Density Cross-correlations

A cross-correlations of the density fields computed using L1 or L2 with the simulation allow a more quantitative appraisal of their respective merits. Figure 13 shows the approximate densities, $1 + \delta_2$ versus those in the simulation, $1 + \delta_1$ at $a = 8$ and $a = 16$. In all cases, the density is obtained by a CIC interpolation to a $64^3$ grid, and concentric lines surround respectively 25, 50, 75, 90 and 99 % of the points. The top row shows the results at $a = 8$. The two other rows correspond to $a = 16$. Both have been smoothed at that time with a gaussian filter of $\ell_s = L_{box}/64$. The middle row corresponds to unsmoothed initial conditions, while the bottom one corresponds to L1S and L2S, i.e. the initial conditions have been smoothed with the same gaussian filter than the one used at the time of comparison.

These plots do confirm the previous findings and show that L2 gives better results than Zeldovich for both expansion factors, in particular for high density contrast $\delta \approx 10$, and with a smaller dispersion around the true values. The correlation coefficients

$$C_1 \equiv \frac{\langle \delta_1 \delta_2 \rangle^2}{\langle \delta_1^2 \rangle \langle \delta_2^2 \rangle}, \quad \text{and} \quad C_2 \equiv \frac{\langle (\delta_1 - \delta_2)^2 \rangle}{\langle \delta_1^2 \rangle},$$

provide global measures of the performance of the approximations. They are given in the upper left corner of each plot and comfort our conclusions. But by very definition, they are mostly sensitive to the densest areas, which hides the fact that L2 is worse than L1 to describe accurately the lowest negative density contrasts.

### 5.4. Count Probability Distribution Function

We now turn to the statistical description of large scale structure in terms of the count probability distribution function (CPDF) and its first few moments. The CPDF allows a simultaneous appraisal of the approximations at different smoothing scales. The CPDF $P_N(\ell)$ is the probability of finding $N$ points in a sphere of radius $\ell$. It is obtained by throwing at random a large number of spheres and by recording there occupation number. As we shall see below, the moments of the CPDF are directly related to the correlation functions $\langle \delta^Q \rangle \equiv \overline{\xi_Q}$ introduced in § 3.

#### 5.4.1. The CPDF

Figure 14 shows the measured $P_N(\ell)$ for different $\ell$ at $a = 16$, both for unsmoothed (top) and smoothed (bottom) initial conditions. The PM result is shown by a solid line, while dots and dashes show those corresponding to L1 and L2, respectively. The measurements scales are, from left to right, $\log_{10} \ell/L_{box} = $ -1.6, -1.4, -1.2, -1.0, -0.8. The corresponding variances (see eq.(85)) measured in the PM simulation are $\langle \delta^2 \rangle \simeq 3.3$, 1.6, 1/1.26, 1/2.5, 1/5; they span all the range of the transition from the weakly to the strongly non-linear regime.

The figure shows that, in the unsmoothed case, L2 corrects L1 in order to make the corresponding approximate CPDF a quasi-perfect description of the PM CPDF, apart from the underdense regime (which are less probable in L2 than in PM, i.e. we recover that L2 underestimates the negative density contrasts), and the densest regime where the improvement brought by L2 over L1 still does not allow to describe the highest and rarest density peaks. In the case of smoothed initial conditions, the L1-CPDF and the L2-CPDF change according to what could be expected from our earlier analysis: L1 now underestimates also the negative density contrasts; the description of weak density contrasts by both L1 amd L2 is somewhat degraded, but smoothing does bring an improve description of the densest part of the simulation. It will prove convenenient to rephrase those statements in terms of values of the number of standard deviation, which we now compute, as well as higher moments.

#### 5.4.2. Moments of the CPDF

The central moments of order $Q$ of $P_N(\ell)$ are defined by

$$\mu_Q(\ell) \equiv <(N - \overline{N})^Q> = \sum_{N=0}^{\infty} (N - \overline{N})^Q P_N(\ell), \qquad (82)$$

with

$$\overline{N} = n\ell^3. \qquad (83)$$

This in turns allow to compute the $Q$-body averaged correlation functions

$$\overline{\xi}_Q(\ell) \equiv v^{-Q} \int_v d^3\mathbf{r}_1 ... \int_v d^3\mathbf{r}_Q \xi_Q(\mathbf{r}_1, ..., \mathbf{r}_Q). \qquad (84)$$

Indeed, $\overline{\xi}_Q(\ell)$ is given for $Q \leq 5$ by (Fry & Peebles 1978, Peebles 1980 for the four first orders)

$$\overline{N}^2 \overline{\xi}_2(\ell) = \mu_2 - \overline{N}, \qquad (85)$$

$$\overline{N}^3 \overline{\xi}_3(\ell) = \mu_3 - 3\mu_2 + 2\overline{N}, \qquad (86)$$

$$\overline{N}^4 \overline{\xi}_4(\ell) = \mu_4 - 6\mu_3 - 3\mu_2^2 + 11\mu_2 - 6\overline{N}, \qquad (87)$$

$$\overline{N}^5 \overline{\xi}_5(\ell) = \mu_5 - 10\mu_4 - 10\mu_2\mu_3 + 35\mu_3 + 30\mu_2^2 - 50\mu_2 + 24\overline{N}. (88)$$

These expressions take the discreteness of the set into account. The larger is $Q$, the larger are the contributing values of $N$ in the sum (82). The high-order correlations are thus related to high density contrasts.

As shown earlier in § 3, in the weakly non-linear regime, the term of order $n - 1$ in the density contrast is required to compute $\langle \xi_n(\ell) \rangle$. And therefore, the term of order $n-1$ in the Lagrangian displacement field is necessary to make a correct estimate of the $S_n$: L1 gives exactly the evolution of the variance in the weakly non-linear regime, and a fraction of the correct answer for the skewness or the kurtosis of the PDF. L2 yields the correct weakly non-linear answer for the skewness.

Figure 15 shows the normalized third, forth and fifth moment of the density distribution measured in the simulation and in the L1 and L2 cases, as functions of the computed variance $\overline{\xi_2}$. The measurement scales $\ell$ lie between $-2.2 \leq \log(\ell/L_{\rm box}) \leq -0.8$. At large scales, we do recover what is expected form the perturbation theory. But we also see that L2 gives reasonably accurate values for $S_5$, which should require in principle a third order calculation. Even more insteresting is the transition regime, at scales $\ell$ such that $\langle \xi_n(\ell) \rangle \sim 1$. Here, even if each approximation under-estimates the second moment at scales progressively more strongly non-linear, the ratios $S_3, S_4, S_5$ remain much better estimated by L2, up to variance $\approx 2$. It was already known that L1 gives, even when $\langle \xi_n(\ell) \rangle \sim 1$ a good qualitative results, even outside of its proper validity range. Figure 15 shows that the "miracle" extends to L2, with the added advantage of giving a much better quantitative description than L1, up to $S_5$. In view of this result, it seems doubtful that third order theory L3 would give appreciably better results than L2.

### 5.4.3. Standard Distribution

Finally, we can use the computed variance (which includes discreteness effects, but they are negligible for all scales but the smallest one shown) $\sigma \equiv \mu_2$ in order to produce normalised distribution function $P_\nu(\sigma)$, with $\nu = (N - \overline{N})/\sigma$. The corresponding curves are shown on figure 16. It shows again that smoothing helps only in the description of rarest events, and is otherwise rather detrimental. A rough measure of the quality of the approximation may be obtained by searching, for a given $\sigma$, what is the largest $\nu$ for which a particular accuracy criterion is met, e.g. that the approximate CPDF does not deviate by more than a factor of two from the PM one. The results of such a measure are given in table 1.

### 5.5. Velocity Field

We now analyze the velocity field, in terms of moduli and angles. Figure 17 shows the modulus of velocity in the two approximations versus the true velocity modulus. The velocity are in units of the Hubble velocity across the box. As before, at $a = 16$, the velocity field was smoothed before its modulus was computed (both for unsmoothed and smoothed initial conditions with a gaussian filter of size $L_{box}/64$). In all cases, L2 provides a substantial improvement over L1, which tend to overestimate the velocities.

| $\sigma^2_{PM}$ | L1 $\nu_M$ | L2 $\nu_M$ | L1S $\nu_M$ | L2S $\nu_M$ |
|---|---|---|---|---|
| 0.2 | 3.9 | 4.4 | 3.9 | 4.4 |
| 0.4 | 4.5 | 6.2 | 4.4 | 6.5 |
| 0.8 | 6.7 | 7.2 | 6.7 | 11.3 |
| 1.6 | 9.1 | 10.4 | 9.8 | 12.9 |
| 3.3 | 12.5 | 14.6 | 14.7 | 17.8 |

**Table 1.** The largest number, $\nu_M$, of standard deviations, $\sigma$ for which the approximate CPDF does not deviate by more than a factor of two from the PM one. The first column gives the PM variance, $\sigma^2_{PM}$, the following ones give $\nu_M$ for L1 and L2 applied to unsmoothed initial conditions, and then L1S and L2S, i.e. L1 and L2 applied to smooth initial conditions. These measurements were made at $a = 16$. second and third one give

We can also see that smoothing of the initial conditions, is on this test clearly beneficial.

Let us now define a misalignment angle descriptor,

$$c(\boldsymbol{x}) \equiv \cos(\boldsymbol{V}_{\rm s}(\boldsymbol{x}), \boldsymbol{V}_{\rm a}(\boldsymbol{x})) = \frac{\boldsymbol{V}_{\rm s}(\boldsymbol{x}).\boldsymbol{V}_{\rm a}(\boldsymbol{x})}{V_{\rm s}(\boldsymbol{x}) V_{\rm a}(\boldsymbol{x})}, \qquad (89)$$

where $\boldsymbol{V}_{\rm s}(\boldsymbol{x})$ and $\boldsymbol{V}_{\rm a}(\boldsymbol{x})$ are respectively the simulated and approximated velocity measured at the same point $\boldsymbol{x}$. This function $c$ clearly lies between -1 and 1 and we further define $\mathcal{N}(c) dc$ as the number of points in the interval $[c, c + dc]$. Only $0 \leq c \leq 1$ is showed, but the number of points with negative cosine is negligible and continuously decreases with decreasing $c$.

For two uncorrelated velocity fields $\mathcal{N}(c)$ would be a constant independent of $c$, whereas positively correlated fields $\mathcal{N}(c)$ would lead to a peack at $c = 1$. Figure 18 shows this distribution function for L1 and L2. At $a = 8$, the velocity field given by L2 is a little more correlated with the simulation than the one given by Zeldovich, whereas for $a = 16$ the two approximations give rather similar results. And we find again that at $a = 16$, smoothing of the initial conditions does improve the comparison with the PM results.

The improvement brought by L2 as compared to L1 is less pronounced for this misalignment angle than for the velocity moduli. But overall L2 gives a much better estimation of the velocity field than Zeldovich approximation. Zeldovich approximation gives good results for the density evolution mainly because, as any consistent Lagrangian approach, it conserves mass and momentum at first order; the good L2 results shows the improvement brought by conserving them at second order.

### 6. Conclusions

In this paper, we extended further the perturbative Lagrangian approach initiated by Moutarde et al. (1991). We gave in particular the growth rates (for the fastest growing modes) up to third order for the matter era of Friedman-Lemaître models with arbitrary density parameter $\Omega$, with either a zero cosmological constant, $\Lambda = 0$, or in the spatially flat case, $\Omega + \Lambda = 1$.

We used these results to compute the skewness of the density field, $\langle \delta^3 \rangle$, in the weakly non-linear regime, both for gaussian and a broad class of non-gaussian initial conditions. We then showed how to compute the same quantity in redshift space, when the "particle" Eulerian positions, $\boldsymbol{r}$ are estimated by $\dot{\boldsymbol{r}}/H$. It shows that the relation between skewness and variance, which is unaffected by this transformation in the gaussian case, may be quite modified for non-gaussian initial conditions, depending on the strength of the initial deviations from "gaussianity".

We then checked whether the second order correction to Zeldovich (linear) approximation improved the well known ability of that approximation to follow some aspects of the dynamics into the non-linear regime, even when the validity conditions of these approximations are not fulfilled anymore. We found that this is indeed the case, both for various spherically symmetric perturbations of known evolution and by detailed comparisons with a numerical simulation starting from gaussian initial conditions of spectral index $n = -2$.

The main improvements brought by the second order correction concern the description of regions with large density contrasts ($\delta \gg 1$) and of the velocity field. We found that the density PDF is computed with reasonable accuracy up to as much as 18 standard deviations, when the latter reaches about 1.8. And indeed the five first order moments of the PDF are accurately estimated till that stage, while Zeldovich approximation would quite underestimate them. The improvements brought to the description of the velocity field are also rather spectacular, as may be judged for instance by the cross-correlation coefficients $C_1$ and $C_2$.

Given that the second order correction to Zeldovich approximation retains the advantages of the latter, i.e its ability to derive any stage of the evolution by simple calculations performed at an initial stage, we conclude that this second order Lagrangian approximation will be quite useful for the understanding of large scale structures dynamics under gravity.

*Acknowledgements.* SC is supported by DOE and by NASA through grant NAGW-2381 at Fermilab. Part of this work was done while SC was at the Institut d'Astrophysique de Paris (CNRS), supported by Ecole Polytechnique. RJ thanks Alain Omont and John Bahcall for their hospitality respectively at Institut d'Astrophysique de Paris and Institute for Advanced Study, where this research was conducted. The computational means (CRAY-98) were made available thanks to the scientific council of the Institut du Développement et des Ressources en Informatique Scientifique.

## A. Spherical model: solutions in various approximations

Let us start from an initial profile $\delta_i(x, t_i)$ given at a redshift $z_i$. We compute here various indicators, such as the density contrast, the divergence of the velocity field and the velocity field, at the time $t$, corresponding to present time so to $z = 0$. We assume that $z_i \gg 1$ so that transient terms have disappeared. Then, one can take into account only the faster increasing term. We also recall here the exact solution (Peebles, 1980).

By integrating the total amount of matter inside a sphere (see Martel & Freudling, 1991), mass conservation is written

$$q = x_i[1 + \delta'_i]^{1/3}, \quad x_i \equiv x(q, t_i), \tag{A1}$$

with

$$\delta'_i = \frac{3}{x_i^3} \int_0^{x_i} y^2 \delta_i(y) dy. \tag{A2}$$

Thus, the first order displacement $\tilde{\Psi}^{(1)}$ is

$$\tilde{\Psi}^{(1)}(q) = x_i\{1 - [1 + \delta'_i]^{1/3}\} \tag{A3}$$

The Lagrangian density contrast is written

$$\delta = \frac{q^2}{x^2}\frac{dq}{dx} - 1, \tag{A4}$$

with $x = q + (g_1/g_{1,i})\tilde{\Psi}^{(1)} + (g_2/g_{1,i}^2)\tilde{\Psi}^{(2)} + \mathcal{O}(\varepsilon^3)$. The calculation of $\tilde{\Psi}^{(2)}$ easily leads to

$$\tilde{\Psi}^{(2)} = \left[\tilde{\Psi}^{(1)}\right]^2 / q. \tag{A5}$$

In the Eulerian case, we naturally obtain

$$\tilde{\Psi}^{(1e)}(x) = \frac{x}{3}\delta'_i, \quad \tilde{\Psi}^{(2e)} = \left[\tilde{\Psi}^{(1e)}\right]^2 / x. \tag{A6}$$

From this comparison to simulation for a initial power spectrum with little power at small scales it seems that second order Lagrangian approximation is a real improvement to first order approximation (Zeldovich) for on many aspects. It can describe higher density contrast, even if it is deficient to describe voids. And its validity after shell-crossing is smaller than with Zeldovich approximation. It gives surprisingly good results for the third, forth and fifth normalized moments of density distribution. L2 gives good results for velocity field, at least for its modulus.

### A.1. Spherical top hat model

#### A.1.1. The density contrast at the center of the perturbation

Here, after having recalled the exact solution (Peebles 1980), we compute, for various perturbative models, the density contrast $\delta$ at the center of the perturbation. $\delta$ does not depend (before shell-crossing) on the considered profile. Spatial derivatives will be done with respect to the comoving coordinate $\mathbf{x}$.

*Exact solution*

Once the value of $\delta_i = \delta(0, t_i)$ is given, the density contrast $\delta$ at time $t$ can be analytically computed whatever the value of the density parameter $\Omega$. We recall here the exact solution (see Peebles 1980).

Let us define the quantities

$$F(\Omega) = \begin{cases} \dfrac{1}{2}\dfrac{1}{1/\Omega - 1} \equiv \dfrac{1}{\mathrm{ch}\,\eta - 1}, & \Omega < 1, \\ \dfrac{3}{10}\dfrac{a_i}{a}, & \Omega = 1, \\ \dfrac{1}{2}\dfrac{1}{1 - 1/\Omega} \equiv \dfrac{1}{1 - \cos\eta}, & \Omega > 1, \end{cases} \quad (A7)$$

$$\delta_c = \frac{3}{5}(1/\Omega_i - 1), \tag{A8}$$

$$\widetilde{\delta}_c = \begin{cases} \delta_c, & \Omega \ne 1 \\ 1, & \Omega = 1. \end{cases} \tag{A9}$$

$\Omega_i$ is the value of the density parameter at $t = t_i$. It is given, as a function of $\Omega$, by the following expression:

$$\Omega_i = \frac{1 + z_i}{1 + \Omega z_i}\Omega. \tag{A10}$$

With these definitions, $\delta(0)$ is

$$\delta_E(0) = \begin{cases} \left|\dfrac{(\delta_c - \delta_i)/\widetilde{\delta}_c}{F(\Omega)(\mathrm{ch}\,\theta - 1)}\right|^3 - 1, & \delta_i < \delta_c, \\ \left|\dfrac{(\delta_c - \delta_i)/\widetilde{\delta}_c}{F(\Omega)(1 - \cos\theta)}\right|^3 - 1, & \delta_i > \delta_c, \end{cases} \tag{A11}$$

where $\theta$ if the "proper conformal time" of the perturbation. It is determined by the implicit equation

$$\begin{cases} \mathrm{sh}\,\theta - \theta = |(\delta_c - \delta_i)/\widetilde{\delta}_c|^{3/2} G(\Omega), & \delta_i < \delta_c, \\ \theta - \sin\theta = |(\delta_c - \delta_i)/\widetilde{\delta}_c|^{3/2} G(\Omega), & \delta_i > \delta_c, \end{cases} \tag{A12}$$

with

$$G(\Omega) = \begin{cases} \mathrm{sh}\,\eta - \eta, & \Omega < 1, \\ \dfrac{4}{3}\left|\dfrac{5}{3}\dfrac{a}{a_i}\right|^{3/2}, & \Omega = 1, \\ \eta - \sin\eta, & \Omega > 1. \end{cases} \tag{A13}$$

*Eulerian linear theory*

Equation (68), taken at first order in $\epsilon$, reads

$$\delta_{E1}(0) = \frac{g_1}{g_{1,i}}\delta_i. \tag{A14}$$

*Eulerian second order perturbation theory*

Equation (68) gives

$$\delta_{E2}(0) = \frac{g_1}{g_{1,i}}\delta_i + \left[\frac{g_1}{g_{1,i}}\right]^2\left(\frac{2}{3} - \frac{1}{3}\frac{g_2}{g_1^2}\right)\delta_i^2. \tag{A15}$$

Therefore, the minimal density contrast that can be described in this approximation is

$$\delta_{\min,E2} = -\frac{3}{4}\left(2 - \frac{g_2}{g_1^2}\right)^{-1}. \tag{A16}$$

*Zeldovich approximation*

Mass conservation is here written as follows

$$\delta_{L1}(0) = \left(1 + \frac{g_1}{g_{1,i}}P\right)^{-3} - 1, \tag{A17}$$

with

$$P \equiv (1 + \delta_i)^{-1/3} - 1. \tag{A18}$$

*Lagrangian second order perturbation theory*

Mass conservation reads

$$\delta_{L2}(0) = \left(1 + \frac{g_1}{g_{1,i}}P + \frac{g_2}{g_{1,i}^2}P^2\right)^{-3} - 1. \tag{A19}$$

The smallest density contrast that can be reached in this approximation then is

$$\delta_{\min,L2} = \left(1 - \frac{1}{4}\frac{g_1^2}{g_2}\right)^{-3} - 1. \tag{A20}$$

A.1.2. *The relationship* $-\nabla.\mathbf{v}/(aH\delta) = f(\delta)$

We assume here that $\delta_i(x, t_i)$ is a continuously differentiable function for any $x \geq 0$. In the vicinity of the perturbation center, the initial density contrast can then be expanded as $\delta_i(x, t_i) = \delta_i(0) + \mathcal{O}(x)$. A sphere of very small radius $x$ is thus similar to a piece of homogeneous universe of density $\overline{\rho}[1 + \delta_i(0)]$. So the proper velocity of an element of matter can be written:

$$u(x) = a_p H_p x + \mathcal{O}(x^2), \tag{A21}$$

where $a_p$ and $H_p$ are respectively the expansion factor and the Hubble constant of this fictitious universe. The peculiar velocity then is

$$v(x) = (a_p H_p - aH)x + \mathcal{O}(x^2). \tag{A22}$$

The quantity we are interested in is written, in the limit $x \to 0$

$$-\frac{\nabla.\mathbf{v}}{aH\delta}(0) = -\frac{3}{\delta}\left(\frac{a_p H_p}{aH} - 1\right). \tag{A23}$$

*Exact solution*

From Eq. (A23), the exact solution is simply (see also Regös et al. 1989)

$$\left[-\frac{\nabla\cdot\mathbf{v}}{aH\delta}\right]_{\mathrm{E}} = \begin{cases} -\dfrac{3}{\delta}\left[\dfrac{\operatorname{sh}\theta(\operatorname{sh}\theta-\theta)}{Ht(\operatorname{ch}\theta-1)^2} - 1\right], & \delta < \delta_{\mathrm{turn}}, \\ -\dfrac{3}{\delta}\left[\dfrac{\sin\theta(\theta-\sin\theta)}{Ht(1-\cos\theta)^2} - 1\right], & \delta > \delta_{\mathrm{turn}}, \end{cases} \quad (A24)$$

with

$$\delta_{\mathrm{turn}} = \begin{cases} \dfrac{2}{9}\dfrac{(\operatorname{ch}\eta-1)^3}{(\operatorname{sh}\eta-\eta)^2} - 1, & \Omega < 1, \\ 0, & \Omega = 1, \\ \dfrac{2}{9}\dfrac{(1-\cos\eta)^3}{(\eta-\sin\eta)^2} - 1, & \Omega > 1. \end{cases} \quad (A25)$$

The quantity $Ht$ can also be written as a function of conformal time $\eta$:

$$Ht = \begin{cases} \dfrac{\operatorname{sh}\eta(\operatorname{sh}\eta-\eta)}{(\operatorname{ch}\eta-1)^2}, & \Omega < 1, \\ \dfrac{2}{3}, & \Omega = 1, \\ \dfrac{\sin\eta(\eta-\sin\eta)}{(1-\cos\eta)^2}, & \Omega > 1. \end{cases} \quad (A26)$$

To obtain the quantity $-\nabla\cdot\mathbf{v}/(aH\delta)$ as a function of $\delta$, we need to suppress $\delta_i$ in the implicit equation (A12). We get

$$\begin{cases} \dfrac{(\operatorname{sh}\theta-\theta)^2}{(\operatorname{ch}\theta-1)^3} = \dfrac{2}{9}\dfrac{1+\delta}{1+\delta_{\mathrm{turn}}}, & \delta < \delta_{\mathrm{turn}}, \\ \dfrac{(\theta-\sin\theta)^2}{(1-\cos\theta)^3} = \dfrac{2}{9}\dfrac{1+\delta}{1+\delta_{\mathrm{turn}}}, & \delta > \delta_{\mathrm{turn}}. \end{cases} \quad (A27)$$

*Eulerian linear theory*

Taking the divergence of expression (71) at first order in $\varepsilon$ reads, using Eq. (A14),

$$\left[-\frac{\nabla\cdot\mathbf{v}}{aH\delta}\right]_{\mathrm{E1}} = f_1, \quad (A28)$$

where $f_1$ is defined in Sect. 2.

*Eulerian second order perturbation theory*

Divergence of expression (71) gives, using Eq. (A15),

$$\left[-\frac{\nabla\cdot\mathbf{v}}{aH\delta}\right]_{\mathrm{E2}} = f_1\frac{\tilde\delta_{\mathrm{E2}}}{\delta} - \left(\frac{g_2}{g_1^2}f_2 - f_1\right)\frac{\tilde\delta_{\mathrm{E2}}^2}{3\delta}, \quad (A29)$$

where $\tilde\delta_{\mathrm{E2}}$ is the solution (if it exists) of the following polynomial

$$\tilde\delta_{\mathrm{E2}} + \left(\frac{2}{3} - \frac{1}{3}\frac{g_2}{g_1^2}\right)\tilde\delta_{\mathrm{E2}}^2 = \delta, \quad (A30)$$

and $f_2$ is defined in Sect. 2.

*Zeldovich approximation*

Equations (A17), (A18) and divergence of the first member of Eq. (70) give

$$\left[-\frac{\nabla\cdot\mathbf{v}}{aH\delta}\right]_{\mathrm{L1}} = -3\frac{(1+\delta)^{-1/3}-1}{\delta(1+\delta)^{-1/3}}f_1. \quad (A31)$$

*Lagrangian second order perturbation theory*

With Eqs. (A19) and (70), we easily obtain

$$\left[-\frac{\nabla\cdot\mathbf{v}}{aH\delta}\right]_{\mathrm{L2}} = -\left(\frac{3}{\delta}\right)\frac{f_1\tilde\delta_{\mathrm{L2}} + \frac{g_2}{g_1^2}f_2\tilde\delta_{\mathrm{L2}}^2}{1+\tilde\delta_{\mathrm{L2}}+\frac{g_2}{g_1^2}\tilde\delta_{\mathrm{L2}}^2}, \quad (A32)$$

where $\tilde\delta_{\mathrm{L2}}$ is the solution (if it exists) of the following polynomial

$$1 + \tilde\delta_{\mathrm{L2}} + \frac{g_2}{g_1^2}\tilde\delta_{\mathrm{L2}}^2 = (1+\delta)^{-1/3}. \quad (A33)$$

### A.2. Velocity field and density contrast for a given initial profile

Here, we try to evaluate the density contrast and the velocity field, as functions of radius, for a given initial profile $\delta_i(x)$. We recall the exact solution and give the various perturbative solutions.

#### A.2.1. The density contrast

*Exact solution*

With the notations we use in this appendix, the exact solution can be written, for $\Omega \neq 1$, for respectively $\delta'_i < \delta_{\mathrm{c}}$ and $\delta'_i > \delta_{\mathrm{c}}$

$$\delta(r) = \begin{cases} \dfrac{|1-\delta'_i/\delta_{\mathrm{c}}|^3\,[F(\Omega)(\operatorname{ch}\theta-1)]^{-3}}{1+3\dfrac{\delta_i-\delta'_i}{\delta_{\mathrm{c}}-\delta'_i}\left[1-\dfrac{3}{2}\dfrac{\operatorname{sh}\theta(\operatorname{sh}\theta-\theta)}{(\operatorname{ch}\theta-1)^2}\right]} - 1, \\ \dfrac{|1-\delta'_i/\delta_{\mathrm{c}}|^3\,[F(\Omega)(1-\cos\theta)]^{-3}}{1+3\dfrac{\delta_i-\delta'_i}{\delta_{\mathrm{c}}-\delta'_i}\left[1-\dfrac{3}{2}\dfrac{\sin\theta(\theta-\sin\theta)}{(1-\cos\theta)^2}\right]} - 1. \end{cases} \quad (A34)$$

For $\Omega = 1$, we have

$$\delta(r) = \begin{cases} \dfrac{9}{2}\dfrac{(\operatorname{sh}\theta-\theta)^2(\operatorname{ch}\theta-1)^{-3}}{1+3(1-\delta_i/\delta'_i)\left[1-\dfrac{3}{2}\dfrac{\operatorname{sh}\theta(\operatorname{sh}\theta-\theta)}{(\operatorname{ch}\theta-1)^2}\right]} - 1, \\ \dfrac{9}{2}\dfrac{(\theta-\sin\theta)^2(1-\cos\theta)^{-3}}{1+3(1-\delta_i/\delta'_i)\left[1-\dfrac{3}{2}\dfrac{\sin\theta(\theta-\sin\theta)}{(1-\cos\theta)^2}\right]} - 1, \end{cases}$$

for $\delta'_i < \delta_c$ and $\delta'_i > \delta_c$ respectively. This calculation is made in Lagrangian coordinates. In other words, the radius $r(t) \equiv ax(t)$ where $\delta$ is evaluated follows the motion. It is related to the initial radius $r_i \equiv a_i x_i$ by

$$r = \begin{cases} \dfrac{3}{10} r_i \dfrac{\operatorname{ch}\theta - 1}{\delta_c - \delta'_i}, & \delta'_i < \delta_c, \\[2ex] \dfrac{3}{10} r_i \dfrac{1 - \cos\theta}{\delta'_i - \delta_c}, & \delta'_i > \delta_c. \end{cases} \quad (A35)$$

The proper conformal time $\theta$ is determined by the implicit equation (A12), in which $\delta_i$ has to be replaced by $\delta'_i$.

*Eulerian linear theory*

The density contrast is simply written

$$\delta_{E1}(x) = \frac{g_1}{g_{1,i}} \delta_i, \quad (A36)$$

*Eulerian second order perturbation theory*

Equation (68) reads

$$\begin{aligned}\delta_{E2}(x) = {} & \frac{g_1}{g_{1,i}} \delta_i \\ & + \left(\frac{g_1}{g_{1,i}}\right)^2 \left(\delta_i^2 + \frac{1}{3}\delta'^2_i - \frac{2}{3}\delta_i \delta'_i + \frac{x}{3}\frac{d\delta_i}{dx}\delta'_i\right) \\ & - \frac{g_2}{g_{1,i}^2}\left(\frac{2}{3}\delta_i \delta'_i - \frac{1}{3}\delta'^2_i\right). \end{aligned} \quad (A37)$$

*Zeldovich approximation*

With the notations given in the beginning of this appendix, we simply have

$$\delta_{L1}(q) = \left(1 + \frac{g_1}{g_{1,i}}\frac{\tilde{\Psi}^{(1)}}{q}\right)^{-2} \left(1 + \frac{g_1}{g_{1,i}}\frac{d\tilde{\Psi}^{(1)}}{dq}\right)^{-1} - 1. (A38)$$

This quantity is evaluated at $x = q + (g_1/g_{1,i})\tilde{\Psi}^{(1)}$.

*Lagrangian second order perturbation theory*

One easily has

$$\begin{aligned}\delta_{L2}(q) = {} & \left(1 + \frac{g_1}{g_{1,i}}\frac{\tilde{\Psi}^{(1)}}{q} + \frac{g_2}{g_{1,i}^2}\frac{\tilde{\Psi}^{(2)}}{q}\right)^{-2} \\ & \left(1 + \frac{g_1}{g_{1,i}}\frac{d\tilde{\Psi}^{(1)}}{dq} + \frac{g_2}{g_{1,i}^2}\frac{d\tilde{\Psi}^{(2)}}{dq}\right)^{-1} - 1.\end{aligned}$$

This quantity is evaluated at $x = q + (g_1/g_{1,i})\tilde{\Psi}^{(1)} + (g_2/g_{1,i}^2)\tilde{\Psi}^{(2)}$.

A.2.2. *The velocity field*

We compute here the normalized velocity $v/(Hr_1)$, where $H$ is the Hubble constant and $r_1$ is the typical radius of the considered fluctuation (so it exactly follows motion).

*Exact solution*

With the notations we use in this appendix, the exact solution can be written

$$\frac{v(r)}{Hr_1} = \begin{cases} \dfrac{r}{r_1} \dfrac{\operatorname{sh}\theta(\operatorname{sh}\theta - \theta)}{Ht(\operatorname{ch}\theta - 1)^2}, & \delta'_i < \delta_c, \\[2ex] \dfrac{r}{r_1} \dfrac{\sin\theta(\theta - \sin\theta)}{Ht(1 - \cos\theta)^2}, & \delta'_i > \delta_c. \end{cases} \quad (A39)$$

This quantity is calculated at the radius $r$ given by Eqs. (A35). $\theta$ is still determined by the implicit equation (A12), but by replacing $\delta_i$ by $\delta'_i$.

*Eulerian linear theory*

At first order, equation (71) reads

$$\left[\frac{v(x)}{Hr_1}\right]_{E1} = -\frac{1}{3}\frac{r}{Hr_1}\frac{\dot{g}_1}{g_{1,i}}\delta'_i, \quad (A40)$$

with $r = ax$.

*Eulerian second order perturbation theory*

At second order, equation (71) reads

$$\begin{aligned}\left[\frac{v(x)}{Hr_1}\right]_{E2} = {} & -\frac{1}{3}\frac{r}{Hr_1}\left[\frac{\dot{g}_1}{g_{1,i}}\delta'_i - \frac{1}{3}\frac{\dot{g}_2}{g_{1,i}^2}\delta'^2_i \right.\\ & \left. + \frac{g_1 \dot{g}_1}{g_{1,i}^2}\left(\delta'_i \delta_i - \frac{2}{3}\delta'^2_i\right)\right],\end{aligned} \quad (A41)$$

with $r = ax$.

*Zeldovich approximation*

Taking Eq. (70) at first order simply leads to

$$\left[\frac{v(q)}{Hr_1}\right]_{L1} = \frac{a}{Hr_1}\frac{\dot{g}_1}{g_{1,i}}\tilde{\Psi}^{(1)}, \quad (A42)$$

This quantity is evaluated at $x = q + (g_1/g_{1,i})\tilde{\Psi}^{(1)}$.

*Lagrangian second order perturbation theory*

At second order, Eq. (70) reads

$$\left[\frac{v(q)}{Hr_1}\right]_{L2} = \frac{a}{Hr_1}\left(\frac{\dot{g}_1}{g_{1,i}}\tilde{\Psi}^{(1)} + \frac{\dot{g}_2}{g_{1,i}^2}\tilde{\Psi}^{(2)}\right) \quad (A43)$$

This quantity is evaluated at $x = q + (g_1/g_{1,i})\tilde{\Psi}^{(1)} + (g_2/g_{1,i}^2)\tilde{\Psi}^{(2)}$.